\documentclass[%
reprint,
showpacs,
showkeys,
bibnotes,
amsmath,amssymb,aps,prb]{revtex4-1}
\usepackage{graphics}
\usepackage{epsfig}
\usepackage{dcolumn}
\usepackage{color}
\usepackage{bm}
\usepackage{subfigure}
\usepackage{soul} 
\usepackage{hyperref}
\usepackage[mathlines]{lineno}
\begin{document}
\title{Electronic structure and magnetic properties of Mn and Fe impurities near GaAs (110) surface}
\author{M.R. Mahani$^{1}$, M. Fhokrul Islam$^{2}$, A. Pertsova$^{1}$ and C.M. Canali$^{1}$}
\affiliation{$^{1}$Department of Physics and Electrical engineering, Linnaeus University, 391 82 Kalmar, Sweden}
\affiliation{$^{2}$Virginia Commonwealth University, Richmond, Virginia 23284, USA}
\date{\today}
\begin{abstract}
Combining density-functional theory calculations and microscopic 
tight-binding 
models, we investigate theoretically 
the electronic and magnetic properties of individual substitutional transition-metal impurities (Mn and Fe) 
 positioned in the vicinity of the (110) surface of GaAs.
For the case of the $[\rm Mn^{2+}]^0$ plus acceptor-hole (h) complex, 
the results of
a tight-binding model including explicitly the impurity $d$-electrons 
are in good agreement with approaches that treat the spin of the impurity
as an effective classical vector. For the case of Fe, where both 
the neutral isoelectronic $[\rm Fe^{3+}]^0$ 
and the ionized $[\rm Fe^{2+}]^-$ states are relevant 
to address scanning tunneling
microscopy (STM) experiments, 
the inclusion of $d$-orbitals
is essential. 
We find that the in-gap electronic structure of Fe impurities 
is significantly modified by surface effects.
For the neutral acceptor state $[{\rm Fe}^{2+}, h]^0$, 
the magnetic-anisotropy dependence on
the impurity sublayer resembles the case of 
$[{\rm Mn}^{2+}, h]^0$. 
In contrast, for $[{\rm Fe}^{3+}]^{0}$ electronic configuration the
magnetic anisotropy behaves differently and it is considerably smaller. 
For this state we predict 
that it is possible to manipulate the Fe moment, e.g. by an external magnetic field, 
 with detectable consequences in the local density of states probed by STM. 
\end{abstract}
\maketitle
\section{\label{sec:level1}INTRODUCTION:\protect}
The study of the spin of individual transition-metal (TM) dopants in 
a semiconductor host is an emergent field known as magnetic solotronics, 
bearing exciting prospects for novel spintronics devices at the atomic 
scale.~\cite{pm_nam11} The development of scanning tunneling microscopy 
(STM) based techniques enabled the investigation of substitutional dopants 
at a semiconductor surface with unprecedented accuracy and 
degree of details.~\cite{yakunin_prl04, shinada_nat05, yazdani_nat06, wiesendanger_MnInAs, gupta_science_2010, garleff_prb_2010,
Fuec_nnt12, pla_nat12} 
The experimental advances have stimulated
theoretical studies of individual 
magnetic impurities in semiconductors, based both on first-principles 
calculations~\cite{zhao_apl04,sarma_prl04,
PhysRevB.70.085411, PhysRevB.75.195335, ebert_prb09,
fi_cmc_prb_2012} 
and microscopic tight-binding (TB) 
models.~\cite{tangflatte_prl04, tangflatte_prb05, timmacd_prb05,Jancu_PRL_08, 
scm_MnGaAs_paper1_prb09, 
scm_MnGaAs_paper2_prb2010, scm_MnGaAs_paper3_prl011, 
PhysRevLett.105.227202,mc_MF_2013} 
Approaches based on the TB models
are particularly convenient to explore the solotronics limit of dilute
impurities in semiconductor hosts.
For the specific case of Mn dopants on the (110) GaAs surface, 
TB models~\cite{tangflatte_prl04, tangflatte_prb05, 
Jancu_PRL_08, scm_MnGaAs_paper1_prb09,
scm_MnGaAs_paper2_prb2010, scm_MnGaAs_paper3_prl011, mc_MF_2013} 
have provided a quantitative description of the 
electronic and magnetic properties of the associated acceptor states.\\ 
Early work by Tang and Flatt{\'{e}}~\cite{tangflatte_prl04} introduced
a TB model for a single substitutional Mn impurity in bulk GaAs. 
Here the electronic structure of the host is described by a $sp$ TB Hamiltonian, while 
the hybridization between the Mn $d$-levels and the $p$-levels 
of the nearest neighbors As atoms is 
modeled as an effective spin-dependent potential. 
This model captures several of the key features 
of the Mn acceptor physics in GaAs found in experiment,
such as the anisotropic shape of the Mn
acceptor wavefunction. 
However, the inclusion of surface effects are expected to be important for 
a direct comparison with STM experiments, 
where accessible impurities 
are positioned in the proximity of the surface.\\ 	
Later experimental~\cite{koenraad_prb08} 
and theoretical~\cite{Jancu_PRL_08, scm_MnGaAs_paper1_prb09} studies 
have indeed demonstrated a strong influence of the surface 
on the properties of the Mn acceptor state. 
In particular, the electronic structure and magnetic properties 
of Mn on the (110) GaAs surface and in near-surface 
layers have been investigated  theoretically
in Ref.~\onlinecite{scm_MnGaAs_paper1_prb09}. 
This work uses a finite-cluster \textit{sp}$^3$ TB model for GaAs, 
with the Mn impurity spin introduced as a classical spin, exchange-coupled 
to the quantum spins of the nearest-neighbor As atoms. 
For a substitutional Mn in bulk GaAs this model
is in good agreement with Ref.~\onlinecite{tangflatte_prl04}. 
For a Mn on the surface the model finds, in agreement with
experiment\cite{yazdani_nat06, gupta_science_2010}, 
a strongly localized and anisotropic mid-gap acceptor state.
Furthermore, the model makes predictions on the dependence of the
acceptor binding energy and magnetic anisotropy energy
on the subsurface layer where the Mn is positioned.
The former result has been later confirmed experimentally.~\cite{garleff_prb_2010}\\
The theoretical approach used both in Ref.~\onlinecite{tangflatte_prl04} 
and Ref.~\onlinecite{scm_MnGaAs_paper1_prb09}
must be viewed as an effective spin model for the Mn spin,
where the Mn $d$-levels have been integrated out 
and the Mn spin is treated as a classical vector. 
The fact that this effective model
makes predictions consistent with experiment, and
it also agrees with TB approaches retaining a microscopic
description of $d$-levels~\cite{Jancu_PRL_08},
is a strong indication that the effective classical-spin approach
is essentially correct for a Mn impurity in GaAs, characterized by a half-filled $d$-shell.
The model does have some limitations, which are recognized and 
understood.~\cite{mc_MF_2013}
\begin{figure}[htp]
\centering
\includegraphics[scale=0.36]{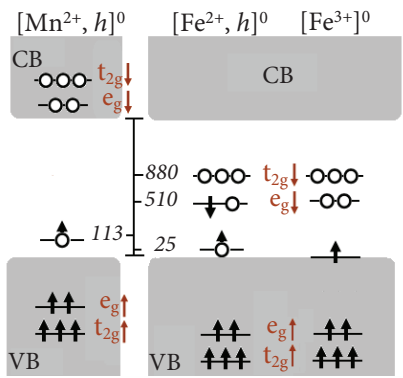}
\caption{Color online -- Transition-metal impurity 
levels in bulk GaAs host. 
The hybridization between Mn $3d$
and nearest neighbor As $3p$ electrons gives rise 
to a bound hole at 113~meV above the valence band
while this weakly bound hole for the case of $[{\rm Fe}^{2+}, h]^0$ is at 25~meV.
In the $[{\rm Fe}^{3+}]^0$ state, the electron is 
filling the weakly bound hole and contributing to the binding.}
\label{fig:elec}
\end{figure}
Other TM impurities in GaAs are less investigated and understood,
and more refined microscopic theories are necessary.
This is one of the motivations of the present work.
The need of a microscopic description of TM impurities 
on a semiconductor surface is also motivated by recent STM experimental studies
of Fe dopants on the (110) surface of GaAs.~\cite{yazdani_prb09,bocq_prb13,svenja_PRB_13} 
Despite similarities in the STM topographies of Mn and Fe impurities 
as a result of the underlying $T_2$ symmetry of the host material~\cite{yazdani_prb09}, the 
electronic structures of  Mn and Fe ions within the GaAs gap 
are known to be 
different Mn~\cite{Malguth_phys_status}, and this should have consequences in the STM
spectroscopic features.
This can already be seen in Fig.~\ref{fig:elec}, where we draw
schematically possible electronic configurations of Mn and Fe ions in bulk GaAs.
In fact, two experimental groups have reported 
somewhat different results in the details of the Fe impurity-induced 
in-gap electronic structure probed by STM.  
Richardella {\it et al.}~\cite{yazdani_prb09} 
found two impurity-induced peaks in their spectroscopic measurements,
while a more recent study reported several peaks    
below and above the valence band maximum, which can be associated 
with the Fe impurity.~\cite{svenja_PRB_13}\\  
Another aspect of the complex electronic structure of Fe dopants on the (110) 
surface of GaAs has been investigated 
in the STM experiments by Bocquel \textit{et al.}~\cite{bocq_prb13} 
The authors demonstrated the possibility of manipulating 
the $d$-shell occupancy of Fe via a voltage-dependent local manipulation 
of the Fermi level due to tip-induced band bending. 
In particular, a transition from an isoelectronic $[{\rm Fe}^{3+}]^0$ 
to an ionized acceptor $[\rm Fe^{2+}]^-$ state ({\it see} Fig.~\ref{fig:elec}) was realized 
by varying the tip voltage. This transition corresponds to the 
change in the filling of the impurity $d$-shell from 
five ($d^{5}$) to six ($d^6$) electrons and therefore to the change 
in the core spin state from $S=5/2$ to $S=2$. 
It is clear that the classical-spin model, used to described 
the neutral $[\rm Mn^{2+}]^0$ plus acceptor-hole (h) complex ($[{\rm Mn}^{2+}, h]^0$), 
is not applicable in this case, as it can not account 
for the change in the valence state of Fe.\\  
In this paper we employ a combined approach using both density functional 
theory (DFT) calculations and microscopic TB modeling to 
study the electronic structure and magnetic properties of substitutional 
Mn and Fe impurities on the (110) GaAs surface and up to 
nine monolayers below the surface. In our TB model we include explicitly 
the $s$-, $p$- and $d$-orbitals of the impurity atom. 
The relative shifts in the on-site energies for the exchange-split minority and 
majority $d$-levels of the dopant are extracted from DFT calculations. 
For the case of Mn we find good agreement 
with previous work, where the Mn magnetic moments was treated classically. This indicates that 
our microscopic TB model correctly reproduces the physics of the $[{\rm Mn}^{2+}, h]^0$  complex,  
while providing access to additional features inaccessible by the classical spin-model.\\ 
In the case of Fe, we perform 
DFT calculations for both bulk and (110) GaAs surface. 
Here our DFT calculations provide a very valuable input for understanding 
the changes in the in-gap electronic states due to the proximity to the surface,   
and for interpreting the results of recent STM experiments. 
Furthermore, using the TB model combined with parameters extracted directly from DFT calculations, 
we provide a detailed analysis of the relevant electronic and magnetic properties of Fe on the (110) GaAs surface. 
In particular, we calculate 
the spatial distribution of the wavefunctions associated with the impurity-induced states 
in the gap as well as the magnetic anisotropy and its  
dependence on the valence state of the impurity. We show that for the neutral acceptor state $[{\rm Fe}^{2+}, h]^0$, 
the magnetic-anisotropy dependence on the impurity position with respect to the surface layer 
is similar to that of $[{\rm Mn}^{2+}, h]^0$. 
In contrast to this, the isoelectronic state, $[{\rm Fe}^{3+}]^0$, 
behaves differently and its anisotropy energy is considerably smaller. This result suggests a route towards manipulation 
of the Fe magnetic moment by external magnetic fields, 
accompanied by detectable changes in the electronic local density of states (LDOS) which can be measured by STM. 
This is different from the case of Mn, 
where such manipulation was shown to be difficult since the large magnetic anisotropy energy (up to 15meV) 
of Mn in the subsurface layers would require 
extremely strong magnetic fields to change the direction of its magnetic moment.~\cite{mc_MF_2013}\\
The rest of the paper is organized as follows. In the next section 
In Sec.~\ref{theo_model}. 
we describe the details of our microscopic  TB approach. 
In Sec.~\ref{DFT} we present the results of DFT calculations for Mn and Fe impurities in GaAs, 
and discuss how these results can be used to extract
some of the parameters of the TB Hamiltonian.
The results of the TB calculations are described in section~\ref{results}.  
Firstly, in section~\ref{Mn dopants} we present the calculations of the electronic energy spectrum, LDOS  
and magnetic anisotropy for the Mn acceptor on the (110) GaAs surface and few layer below it. In particular, we provide a 
quantitative comparison between the results obtained with our fully microscopic TB model 
and the results of the classical spin model for Mn,   
reported previously.~\cite{scm_MnGaAs_paper1_prb09} 
Secondly, in section~\ref{Fe dopants} we address similar properties of Fe dopants both in bulk GaAs and on the (110) 
GaAs surface. We focus in particular on the differences in the in-gap electronic structure, 
LDOS and magnetic anisotropy, associated 
with the Fe impurity in the bulk and on the surface. For the case of Fe on the surface, we show the dependence of the magnetic 
anisotropy on the valence states of the impurity and discuss possible implications on   
STM experiments. Finally, we draw some conclusions.
\section{MICROSCOPIC TIGHT-BINDING MODEL}
\label{theo_model}
We use a multi-orbital TB model to describe TM impurities substituting Ga atoms in GaAs. 
The model includes $s$- and $p$-orbitals for all Ga and As atoms and $s$-, $p$- and $d$-orbitals for 
the impurity atoms. Introducing the $d$-orbitals for the impurities only, and not for Ga atoms,  
is justified by  the fact that, as we will show in the next section,
the $d$-levels of Ga are located far below ($\approx$ 15~eV) the 
Fermi level ({\it see} Fig.~\ref{fig:Mn_Ga_d_DOS}).\\
The second-quantized TB Hamiltonian for (Ga, TM)As takes the following form
\begin{linenomath*}
\begin{align}
H &  =H_{\rm GaAs}+H_{\rm TM} + H_{\rm LRC}%
\label{hamiltonian}
\end{align}
\end{linenomath*}
The first term in 
Eq.~(\ref{hamiltonian}) is the TB Hamiltonian for the GaAs host (with the exclusion of the
Ga atoms replaced by the TM impurity). 
It is the sum of two terms
\begin{equation}
H_{\rm GaAs} =  H_{\rm band} + H_{\rm SOI}\;, 
\end{equation}
where
\begin{equation}
H_{\rm band}   =\sum_{ij,\mu\mu^{\prime},\sigma}t_{\mu\mu^{\prime}}^{ij}a_{i\mu\sigma
}^{\dag}a_{j\mu^{\prime}\sigma}\;,
\label{tb}
\end{equation}
is the usual TB-band Hamiltonian for bulk GaAs\cite{chadi_prb77} written in terms of 
Slater-Koster 
parameters, ($t_{\mu\mu^{\prime}}^{ij}$)~\cite{slaterkoster_pr54,papac_jpcm03}, 
representing both on-site
energies and nearest-neighbors hopping amplitudes. 
Here $a_{i\mu\sigma}^{\dag}$ and $a_{i\mu\sigma}$
are electron creation and annihilation operators; $i$ and $j$ are atomic indices that run over all atoms other than 
the impurity, $\mu$ and $\mu^{\prime}$
are orbital indices and $\sigma = \uparrow, \downarrow $ is a spin index defined with respect to an arbitrary quantization axis. 
The one-body term
\begin{linenomath*}
\begin{align}
H_{\rm SOI} &  =\sum_{i,\mu\mu^{\prime},\sigma\sigma^{\prime}}\lambda_{i}\langle\mu
,\sigma|\vec{L}\cdot\vec{S}|\mu^{\prime},\sigma^{\prime}\rangle a_{i\mu\sigma
}^{\dag}a_{i\mu^{\prime}\sigma^{\prime}}
\label{so}
\end{align}
\end{linenomath*}
models the on-site spin-orbit interaction (SOI) in GaAs, with the values of the  
re-normalized spin-orbit splittings $\lambda_i$ taken from 
Ref.~\onlinecite{chadi_prb77}. 
The second term
in Eq.~(\ref{hamiltonian}) describes the TM impurity and its coupling to the atoms of the host.
We have
\begin{linenomath*}
\begin{align}
H_{\rm TM} &  =\sum_{i,m,\mu,\nu,\sigma}\big( t_{\mu\nu}^{im}a_{i\mu\sigma
}^{\dag}a_{m\nu\sigma} + 
t_{\mu\nu}^{im \star}a_{m\nu\sigma}^{\dag} a_{i\mu\sigma
}\big)
\nonumber\\
&  +\sum_{m,\nu,\sigma}\epsilon_{m \nu \sigma}
a_{m\nu\sigma}^{\dag}a_{m\nu\sigma}\nonumber\\
& + H^{\rm TM}_{\rm SOI}\;,
\label{imp}
\end{align}
\end{linenomath*}
where $a_{m\nu\sigma}^{\dag}$ and $a_{m\nu\sigma}$ are creation and annihilation operators
at the impurity site $m$; the orbital index $\nu$ runs over $s$-, $p$-, and $d$-orbitals.
The first term in 
Eq.~(\ref{imp}) describes a hopping between the impurity and its As nearest-neighbors. 
For the Slater-Koster hopping parameters between the impurity $d$-orbitals and the nearest-neighbor 
As $s$- and $p$-orbitals we use the same values as for the corresponding hopping parameters 
between Ga and As, which are available in the literature.~\cite{Bassani_PRB_57_6493}
Our tight-binding model must be viewed as a phenomenological 
approach which allows us to introduce, in a physically meaningful way, 
the microscopic d-levels for the impurity and to go beyond the classical-spin model used previously.\\
The second term in Eq.~(\ref{imp}) represents on-site energies of the impurity for a given orbital.
The $d$-orbital energies $\epsilon_{m\,d\,\sigma}$ play an important role in the model. 
Their values for ``spin-up'' (majority)
and ``spin-down'' (minority) electrons
are different, which leads to a different occupation for opposite spin states, 
and hence to a non-zero spin magnetic moment at the impurity site.
As a first estimate of the on-site $d$-orbital energies, we use the values of the exchange-split 
majority and minority $d$-levels, which can be identified 
in the spin- and orbital-resolved density of states (DOS) of the impurity, calculated with DFT. 
The exact procedure and the choice of 
the $d$-orbital on-site energies for specific cases of Mn and Fe in GaAs will be discussed in 
section~\ref{DFT}. 
The last term in Eq.~(\ref{imp}) is an on-site spin-orbit coupling term for the impurity atom,
analogous to the one given in  Eq.~(\ref{so}).
The spin-orbit coupling terms $ H_{\rm SOI}$ and $ H^{\rm TM}_{\rm SOI}$ will cause the total 
ground-state energy of the system 
to depend on the direction of the impurity magnetic moment, 
defined with respect to an arbitrary quantization axis. This is the origin of the magnetic anisotropy energy.\\  
Finally, the third term in Eq.~(\ref{hamiltonian}) 
\begin{linenomath*}
\begin{align}
H_{\rm LRC} &  =\frac{e^{2}}{4\pi\varepsilon_{0}\varepsilon_{r}}\sum_{m}\sum_{i\mu\sigma
}\frac{a_{i\mu\sigma}^{\dag}a_{i\mu\sigma}}{|\vec{r}_{i}\mathbf{-}\vec{R}%
_{m}|}\;,%
\label{lrc}
\end{align}
\end{linenomath*}
is a long-range repulsive Coulomb 
potential that is dielectrically screened by the host material
(the index $m$ runs over all impurity atoms). This term prevents extra electrons from approaching 
the impurity atom too
closely and therefore, prevents it from charging. 
This contributes to localize the acceptor hole around the impurity.
The dielectric constant $\epsilon_r$ for the impurity on the GaAs 
surface is reduced from the bulk GaAs value in order to mimic 
a weaker dielectric screening at the surface (12 for bulk and 6 for the surface). 
This crude choice is qualitatively supported by experimental results.\cite{garleff_prb_2010}\\
As already mentioned in the introduction, the modeling of the TM impurity considered here ({\it see} Eq.~\ref{imp})
differs from the approach of Ref.~\onlinecite{scm_MnGaAs_paper1_prb09} 
in that the TM impurity $d$-orbitals are introduced explicitly.
In Ref.~\onlinecite{scm_MnGaAs_paper1_prb09}
the $d$-orbitals are integrated out and their spin-dependent hybridization with the As  $p-$orbitals is represented by 
the following effective spin Hamiltonian
\begin{linenomath*}
\begin{equation}
H_{\rm class-spin} =
J_{pd}\sum_{m}\sum_{i[m]}\vec{S}_{i}\cdot
\hat{\Omega}_{m}\
\label{class_spin}
\end{equation}
\end{linenomath*}
Eq.~\ref{class_spin} describes the antiferromagnetic exchange coupling (with coupling constant $J_{pd} >0 $) 
between the TM impurity spin $\hat{\Omega}_{m}$ 
(treated as a classical vector) and the nearest neighbor
As $p$-spins $\vec{S}_i =  1/2\sum_{\pi\sigma{\sigma'}} a_{i\pi\sigma}^\dagger \vec{\tau}_{\sigma{\sigma'}} a_{i\pi{\sigma'}}$.
Below we will refer to the model of Eq.~\ref{class_spin} as \textit{classical-spin model}, 
in contrast to the present impurity model,
given in Eq.~\ref{imp}, which will be denoted as {\it quantum $d$-level model}.\\
The electronic structure of GaAs with a single substitutional Mn or Fe impurity atom 
is obtained by performing supercell-type calculations with a cubic cluster of 3200 atoms
and periodic boundary conditions in either 2 or 3 dimensions, depending on
whether we are studying the $\left(  110\right)  $ surface or a bulk-like
system. The $\left(  110\right)  $ surface of GaAs is attractive from both
theoretical and experimental points of view due to the absence of large surface 
reconstruction. In order to remove 
artificial dangling-bond states that would otherwise appear in the
band gap, we include relaxation of surface layer positions 
following a procedure put forward in 
Refs.~\onlinecite{chadi_prl78} and~\onlinecite{chadi_prb79}.
For more details the reader is
referred to Ref.~\onlinecite{scm_MnGaAs_paper1_prb09}.
We would like to note here that while the effects of 
strain induced by the (110) surface relaxation in GaAs are 
included in our study, following the prescription of 
Refs.~\onlinecite{chadi_prl78} and~\onlinecite{chadi_prb79} mentioned above, we do not take into 
account the modification in strain and relaxation caused 
by the presence of the magnetic impurity. A systematic study 
of surface-induced strain and strain from an embedded quantum 
dot on the symmetry of an acceptor is presented, respectively, 
in Refs.~\onlinecite{celebi_prl10} and~\onlinecite{paul_nature_2007}.
In Ref.~\onlinecite{PhysRevB.79.245201} the authors
present a model for the on-site matrix elements of the $sp^3d^5s^*$ TB Hamiltonian
of a strained diamond or zinc-blend crystal.
Finally, in a more recent paper~\cite{PhysRevB.86.115424}, the author
presents a method to include strain into the tight-binding Hamiltonian, which
is suitable for calculations of the electronic
and optical properties of semiconductor nanosystems
embedded in a host crystal. These extensions of the TB model represent future challenges to further improve the description of the strain effects induced by surfaces and impurities in semiconductors.
\section{DFT calculations and extraction of the TB parameters}
\label{DFT}
The DFT calculations are performed using the full-potential all-electron method with 
the basis consisting of linearized augmented plane waves combined with local orbitals (LAPW+lo), 
as implemented 
in the WIEN2k package.~\cite{Wien2k_package} We use the generalized gradient approximation (GGA) with 
Perdew-Burke-Ernzerhof (PBE) exchange correlation functional\cite{perdew96} as well as GGA+U, where the 
orbital dependent $U$ parameter is used to capture electronic correlations 
in the core shell of the impurity. 
Because of the time-expensive nature of the plane wave method, we have used the SIESTA \textit{ab initio}
 package, which employs pseudopotentials and a numerical basis set~\cite{siesta}, for relaxing the 
surfaces in our calculations. The relaxed coordinates are then used as input for WIEN2k calculations.\\
For bulk calculations, we consider a $2\times2\times2$ supercell containing a 
total of 64 atoms, with one Ga atom replaced by a TM impurity atom. We use 100 non-equivalent 
$k$-points 
in the Brillouin zone. For the surface calculations, a $4 \times 2$ surface supercell with six layers, 
each containing 16 atoms, is constructed by cleaving the bulk crystal along the $[110]$ direction. A vacuum 
of 25 Bohr is added along the surface to avoid supercell interaction. All the surface calculations are
performed by substituting one Ga atom from the surface layer by a TM impurity. Note that due to computational 
limitations caused by the large size of the supercell we use one k-point in the surface calculations. This 
choice of k-point sampling can be justified based on the fact that the Brillouin zone for the surface 
supercell is considerably smaller than that of the bulk.~\cite{fi_cmc_prb_2012}\\ 
The DFT calculations carried out in this work 
provide the spin-resolved DOS for the $d$-electrons 
of Mn and Fe impurities in the bulk and on the surface of GaAs. Importantly, 
the splitting between the majority and the minority $d$-levels of the impurities, calculated with DFT,  
determines the relative values of the on-site energies for the corresponding 
majority and minority $d$-orbitals  
in our TB model.
\begin{figure}[htp]
\centering
\includegraphics[scale=0.46]{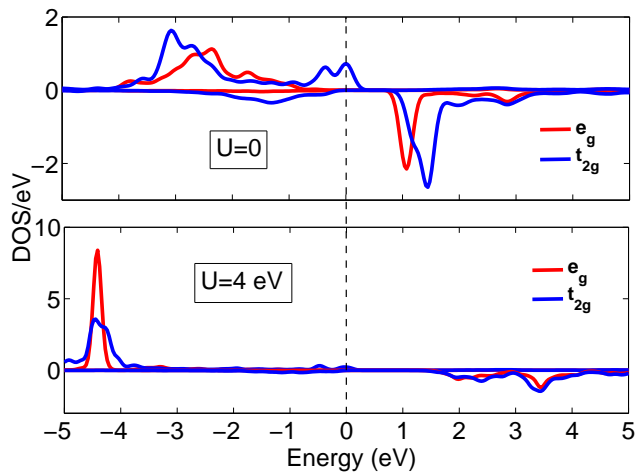}
\caption{Color online -- Spin-resolved DOS for the $d$-orbitals of the Mn dopant in bulk GaAs
for $U=0$ (top panel) and $U=4$~eV (bottom panel). 
The vertical dashed line at $E= 0$ denotes the position of the Fermi level.}
\label{fig:Mn_DOS-B}
\end{figure}
Fig.~\ref{fig:Mn_DOS-B} shows the spin-resolved DOS for the  
Mn dopant in bulk GaAs for two different choices of the $U$ parameter in the GGA+U calculations. 
The $U$ parameter tends to localize the majority $d$-shell electrons 
and push them deeper in the valence band. 
It is clear from the figure that the separation between 
majority (spin-up) and minority (spin-down) $d$-orbitals of different symmetry ($e_g$ and $t_{2g}$) 
increases with increasing the value of $U$.\\ 
A microscopic determination of the parameter U 
in first-principles GGA+U (or LDA+U) calculations in complex 
materials is an outstanding and complicated issue in computational 
physics and material science, details of which were first 
addressed in Ref.~\onlinecite{anisimov1991band}. For the specific 
case of TM impurities in semiconductors, more information is 
available on the value of U for Mn in GaAs via comparison with 
experiment and calculations; while for the 
case of Fe dopants, information is scarcer.  
We believe that neglecting U altogether is 
not justified, since some kind of self-interaction correction 
to the TM must be included and it does have an effect on the 
electronic structure, noticeably the position of the d-levels. 
Our approach has been to choose a value not very different from 
the value of U for Mn, under the assumption that the two cases should be 
qualitatively similar and that small variations of U, if present, 
should not change the results substantially. 
Refs.~\onlinecite{uppsala_DMSreview_2010} and~\onlinecite{belhadji2007trends} 
both use $U=4$~eV in Fe-doped GaN semiconductors. 
In Ref.~\onlinecite{gopal2006magnetic} which investigates several transition-metal-doped ZnO semiconductors 
with LDA+U, the value of $U=4.5$~eV (and exchange $J=0.5$~eV) is used 
for all transition metals. This paper mentions explicitly that, 
although small variations are expected across the TM series, the 
choice of constant values permits a more straightforward comparison, 
which is precisely our point of view on this issue.
The value of $U$ is usually 
chosen to match photoemission spectra
and in our calculation we use $U=4$~eV~\cite{uppsala_DMSreview_2010},
although the results for $U=0$ are also presented.\\
The values of the on-site TB parameters for the Mn $d$-orbitals in bulk 
 are determined based on the calculations with $U=4$~eV ({\it see} the bottom panel of Fig.~\ref{fig:Mn_DOS-B}). 
As a first estimate for these parameters we take the positions of the spin-resolved $e_g$ and $t_{2g}$ levels.  
 We then tune the value of the on-site energy for the majority $t_{2g}$ orbital to get 
the exact position of the acceptor level introduced by Mn in the bulk GaAs gap. 
The values of the on-site energies are summarized in Table~\ref{tab:Mn_param}.\\ 
The spin-resolved DOS for Mn on the (110) GaAs surface is shown in Fig.~\ref{fig:Mn-DOS-S}. 
The lower symmetry on the surface lifts the degeneracy of the $d$-orbitals and,
strictly speaking, we can no longer identify peaks corresponding to orbitals with $e_g$ and $t_{2g}$ 
symmetry as in the bulk case. However, as one can see from the bottom panel 
of Fig.~\ref{fig:Mn-DOS-S}, in the case of $U=4$~eV the positions of the main peaks in 
the surface DOS are quantitatively similar to the positions of the $e_g$ and $t_{2g}$ peaks 
in the bulk calculation with the same value of U (bottom panel of Fig.~\ref{fig:Mn_DOS-B}). 
Therefore,  in our TB calculations for Mn on the (110) GaAs 
surface we use the same set of parameters as for Mn in bulk. 
As we will show in section~\ref{Mn dopants}, 
the properties of the acceptor state obtained from these 
calculations are in good agreement with the STM experiments.\\
\begin{linenomath*}
\begin{table}
\caption{\label{tab:Mn_param}The on-site energies of Mn and Fe $d$ orbitals in eV.}
\begin{ruledtabular}
\begin{tabular}{ccccc}
 & Mn (bulk) & Mn (surface) & Fe (bulk) & Fe (surface)\\
\hline
$e_g^{up}$    &-4.5    &-4.5    &-6.5    &-4.5 \\
$t_{2g}^{up}$ &-2.226  &-2.226  &-6.5    &-4.5 \\
$e_g^{dn}$    & 3.5    & 3.5    &-0.138  &-1   \\
$t_{2g}^{dn}$ & 3.5    & 3.5    &-0.02   &-1   \\
\end{tabular}
\end{ruledtabular}
\end{table}
\end{linenomath*}
\begin{figure}[htp]
\centering
\includegraphics[scale=0.46]{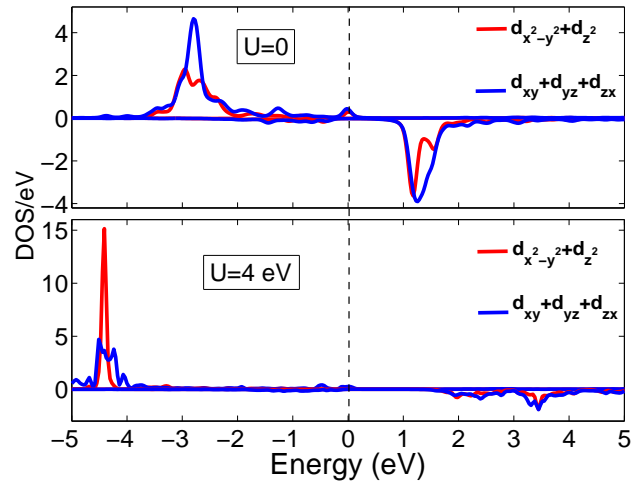}
\caption{Color online -- Spin-resolved DOS for the $d$-orbitals of the Mn dopant 
on the (110) GaAs surface 
for $U=0$ (top panel) and $U=4$~eV (bottom panel). The vertical dashed line at $E=0$ 
denotes the position of the Fermi energy.}
\label{fig:Mn-DOS-S}
\end{figure}
In Fig.~\ref{fig:Mn_Ga_d_DOS} we also plot the spin-resolved density of states 
for the $d$-orbitals of a Ga 
atom on the (110) GaAs surface, and compare it with the analogous quantity for a substitutional
Mn atom. We can clearly see that the Ga $d$-orbitals are, as expected, unpolarized and
occur at energies far below the Fermi level.
\begin{figure}[htp]
\centering
\includegraphics[scale=0.42]{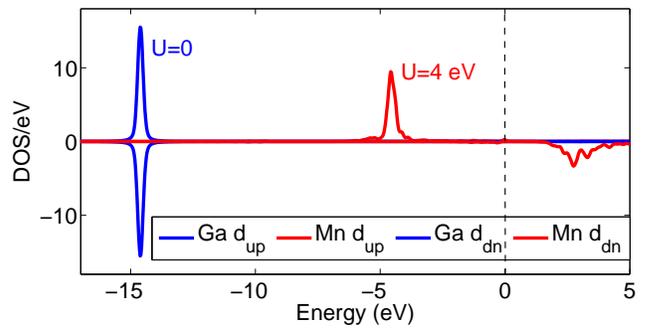}
\caption{Color online -- Spin-resolved density of states (DOS) for the 
$d$-orbitals of a Ga atom and that of a substitutional Mn on the (110) GaAs surface.  
 The vertical dashed line at $E= 0$ denotes the position of the Fermi level.}
\label{fig:Mn_Ga_d_DOS}
\end{figure}
The band structure of bulk GaAs ({\it see} Fig.~\ref{fig:GaAs_band}) 
reveals that the top of the valence band in GaAs is completely 
dominated by $p-$like states, both for Ga and As. Levels of 
$d-$character are visibly present, particularly for Ga 
atoms, even at the Gamma point. However, their contribution 
is only 2\% of the $p-$like states. These results justify our simplifying choice of
disregarding Ga $d$-orbitals in the TB model.
We would like to emphasize that the $d-$level parameters in the empirical 
TB model can differ considerably from the corresponding electronic atomic 
orbitals. Therefore our remark on the position of the \textit{ab-initio} Ga 
$d-$level energy being far away from the Fermi energy as a 
justification of our neglect of these levels in the TB model 
must be taken with caution. Note, however, that the parametrization 
of GaAs that we adopt here, which excludes $d-$levels of Ga and As atoms, is a 
standard procedure and therefore the \textit{ab-initio} results are not inconsistent 
with neglecting the Ga and As $d-$levels in the TB model.\\
\begin{figure}[htp]
\centering
\includegraphics[scale=0.7]{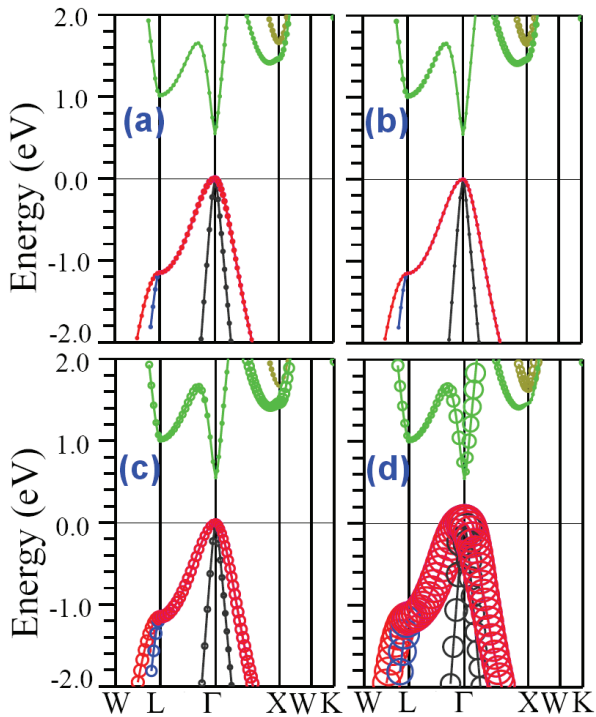}
\caption{Color online -- Band structure of bulk GaAs along the high 
symmetry point in the Brillouin zone. (a)-(d) Band character of Ga $d-$,
As $d-$, Ga $p-$ and As $p-$orbitals respectively. The radius of the circles represents
the corresponding weight for the particular k and energy point.
The horizontal black line at $E=0$ denotes the position of the Fermi level.}
\label{fig:GaAs_band}
\end{figure}
We will now discuss the case of Fe in GaAs. The spin-resolved DOS for the Fe dopant in 
bulk GaAs is presented in Fig.~\ref{fig:DFT-Fe-B}. We note that our calculations 
with $U=0$ for both Fe and Mn in bulk GaAs are in agreement with previously reported results.~\cite{sarma_prl04}
As one can see from Fig.~\ref{fig:DFT-Fe-B},  
 the majority $e_g$ and $t_{2g}$ peaks are pushed deeper in the valence band 
 for $U=4$~eV compared to the case with $U=0$. 
 Interestingly, the positions of the minority $e_g$ and $t_{2g}$ peaks above the Fermi level are sensitive to the value of U. 
The minority doublet ($e_g$) level is lower in energy than the minority triplet ($t_{2g}$) 
level for the case with $U=0$, however these two levels swap for $U=4$~eV. 
Based on the calculation with $U=4$~eV, we choose the positions of the 
two main peaks for the majority $d$-levels (-6.5~eV) as the on-site energy of the corresponding $d$-orbitals 
in our TB model for Fe in bulk.  
We then tune the value of the on-site energy for the minority $e_g$ and $t_{2g}$ orbitals to get 
the exact electronic structure of Fe in bulk GaAs.~\cite{Malguth_phys_status} We tune these two parameters to avoid any complications 
caused by swapping of the two minority peaks for different $U$ parameters.\\
The results of the DFT calculations for the Fe dopant on the (110) surface of GaAs are shown in  
Fig.~\ref{fig:DFT-Fe-S}. The lower symmetry on the surface lifts the degeneracy of the $d$-orbitals and more peaks 
appear in the DOS compared to the bulk. 
The values of the TB parameters of the $d$-orbitals of Fe on the surface are again extracted directly from 
 the DFT calculation with $U=4$~eV ({\it see} the bottom panel of Fig.~\ref{fig:DFT-Fe-S}) as the positions of the peaks in the 
spin-resolved DOS for the corresponding orbitals. 
In contrast to Mn, we do not tune the extracted values 
for Fe on the GaAs surface. This is due to the lack of accepted experimental data on 
 the position of the impurity-induced states in the gap for the case of Fe on the surface.   
We take the value for the spin-up on-site energy of $e_g$ and $t_{2g}$ orbitals from the main peak of the majority DOS, 
which is located at $\approx$4.5~eV. In the case of the minority DOS, we find two pronounced peaks, one above and the  
other one below zero. The choice of the 
positive on-site energy for the spin-down $e_g$ and $t_{2g}$ orbitals  
pushes the Fe levels into the conduction band, which seems to be an 
unlikely scenario based on the bulk level-structure. Therefore 
we take the value for the on-site energy for spin-down $e_g$ and $t_{2g}$ orbitals   
from the peak in the minority DOS located at $\approx$-1~eV ({\it see} Table~\ref{tab:Mn_param}).
\begin{figure}[htp]
\centering
\includegraphics[scale=0.46]{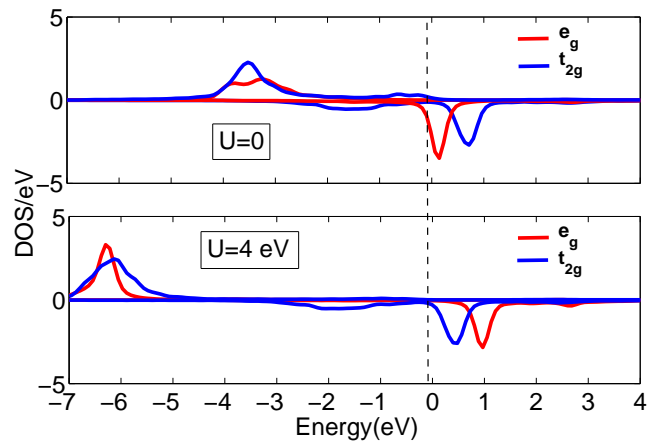}
\caption{Color online -- Spin-resolved DOS for the $d$-orbitals of the Fe dopant in bulk GaAs
for $U=0$ (top panel) and $U=4$~eV (bottom panel). The vertical dashed line 
at $E=0$  denotes the position of the Fermi level.}
\label{fig:DFT-Fe-B}
\end{figure}
\begin{figure}[htp]
\centering
\includegraphics[scale=0.46]{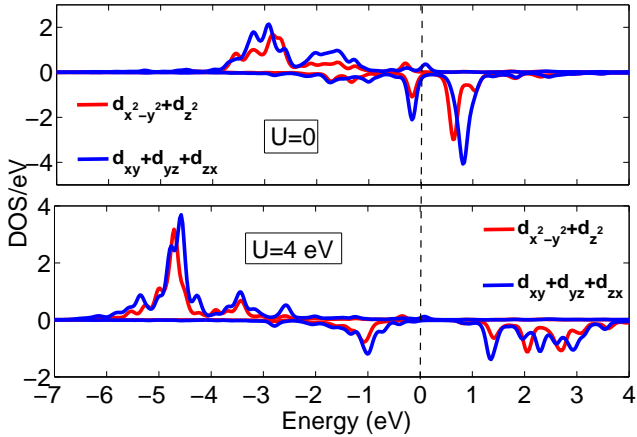}
\caption{Color online -- Spin-resolved DOS for the $d$-orbitals of the Fe dopant 
on the (110) GaAs surface
for $U=0$ (top panel) and $U=4$~eV (bottom panel). 
The vertical dashed line at $E=0$ denotes the position of the Fermi level.}
\label{fig:DFT-Fe-S}
\end{figure}
\section{RESULTS AND DISCUSSION}\label{results}
\subsection{Mn dopants on (110) GaAs surface}
\label{Mn dopants}
\begin{figure}[htp]
\centering
\includegraphics[scale=0.2]{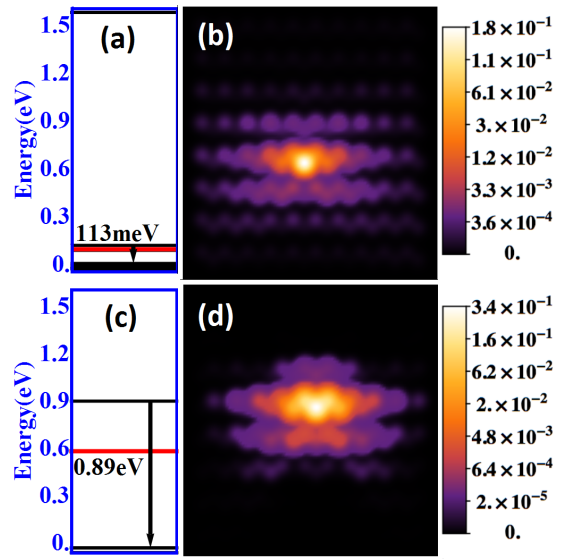}
\caption{Color online -- Electronic properties of Mn in GaAs, calculated using the TB model in which 
the Mn $d$-levels are included explicitly. Eigenvalue spectrum (a,c) and the calculated LDOS for the acceptor state 
(b,d). Top panels are for Mn in bulk, bottom panels are for Mn 
on the surface. In panels (a) and (c) the red lines mark the highest occupied level while 
the black lines mark the acceptor state (the first level above the highest 
occupied level).}
\label{fig:spec-LDOS}
\end{figure}

\begin{figure}[htp]
\centering
\includegraphics[scale=0.2]{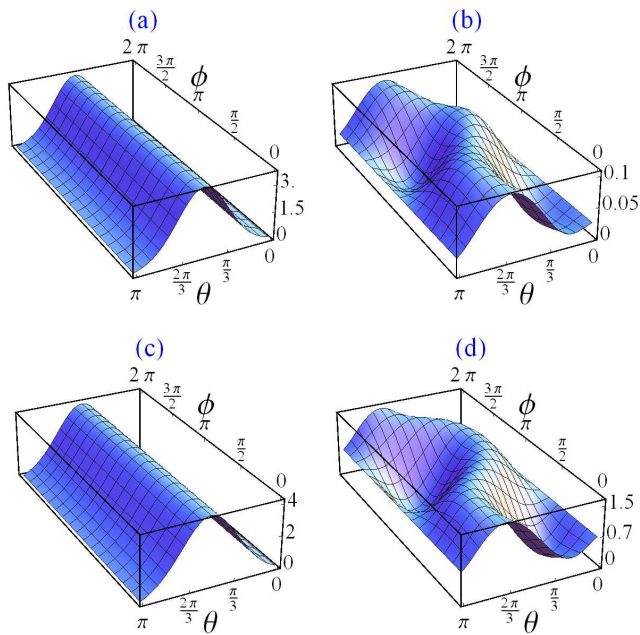}
\caption{Color online -- A comparison of magnetic anisotropy energy landscapes calculated with  
 the fully microscopic model, which includes the Mn $d$-orbitals, (top panels) and with 
the classical-spin model (bottom panels). Panels (a,c) are for bulk, (b,d) are for the surface. 
Energy is in the units of meV.}
\label{fig:MAE_BS}
\end{figure}

We start by discussing the results of TB modeling of a single Mn dopant in GaAs.  
As we will see in the following, our TB model incorporating the impurity $d$-orbitals    
reproduces the well-known features of Mn in the bulk and on the surface of GaAs, 
in agreement with previous 
studies. However,  
we also find some interesting differences between our approach 
and the classical-spin model.~\cite{scm_MnGaAs_paper1_prb09} 
These differences are mainly 
related to the magnetic anisotropy of Mn on the surface and in the subsurface layers.\\ 
Fig.~\ref{fig:spec-LDOS} shows the electronic properties of Mn in the bulk 
(top panels) and on the (110) surface 
(bottom panels) of GaAs. Mn introduces three levels in the GaAs gap, 
with the highest level, which is unoccupied, 
known as the hole-acceptor level. The other two levels are occupied and they lie below the acceptor. 
The position of the acceptor level with respect to the valence band is found 
at 113~meV for the bulk and at 0.89~eV for the surface dopant. 
While the bulk calculation reproduces exactly the 
experimental value~\cite{schairer_prb74, lee_ssc64, chapman_prl67, linnarsson_prb97} 
(see the discussion about the TB parametrization in section~\ref{DFT}), 
the surface calculation also gives 
the position of the acceptor level close to the experimental result.~\cite{yazdani_nat06}
As one can see from Fig.~\ref{fig:spec-LDOS}(c), the calculated LDOS for the acceptor on the surface  
shows more concentration of the spectral weight on the Mn site compared to the bulk case, which is an    
indication of a deeper and a more localized state. 
In general, the calculations presented in Fig.~\ref{fig:spec-LDOS} support the results 
of the classical spin model~\cite{tangflatte_prl04, scm_MnGaAs_paper1_prb09} 
and are in good agreement with other theoretical and 
experimental results.~\cite{Jancu_PRL_08, yazdani_nat06}\\
We would like to comment on one important feature of the 
electronic structure of Mn in bulk GaAs.  
According to Fig.~\ref{fig:spec-LDOS}(a) 
the three levels introduced by Mn in the bulk GaAs gap 
are found to be spread over an energy interval of approximately 30~meV, 
when SOI is included in the calculations. 
(In the Figure, the top-most level and the lowest level in the gap are split
by $\approx 30$ meV).
This is a shortcoming that the present quantum $d$-level model shares with 
the classical-spin models of 
Refs.~\onlinecite{tangflatte_prl04, scm_MnGaAs_paper1_prb09}: In fact
the three (predominantly)
$p-$levels appearing in the gap should be degenerate in the perfectly
tetragonal environment of an impurity in bulk GaAs. The lifting of the
degeneracy is connected with the breaking of time reversal and rotational symmetry
in mean-field-like treatments of the kinetic-exchange coupling between
the TM impurity $d$-levels and the $p-$levels of the nearest neighbor As atoms.\\
The symmetry of the Mn acceptor ground state in bulk GaAs 
is an important issue. The correct framework to discuss this 
problem is within a many-body approach. For a Mn in bulk GaAs, 
a simplified many-body Hamiltonian that captures the salient 
features of the problem consists of the Mn impurity (with ten 
$d-$levels occupied on average by five electrons) and the four 
nearest-neighbor As atoms. The problem is eventually reduced 
to a sum of three terms: the acceptor Hamiltonian (with 
single-particle degeneracy equal to six corresponding to 
twice the number of $p-$levels) which in the GS is spanned by 
states with five electrons (or, equivalently, one hole); the 
Mn-impurity part with (on average) five electrons localized 
in the $d-$orbitals of Mn$^{2+}$; and a hopping Hamiltonian describing 
the $p-d$ hybridization between As $p-$orbitals and Mn$^{2+}$ $d-$orbitals. 
This finite-cluster many-electron model captures the tetragonal 
symmetry of the system; although (being a finite system), it is 
only a rudimentary representation of the valence-band structure 
at the $\Gamma$ point.\\ 
The many-body Hamiltonian can be solved 
approximately by second order perturbation theory in the hopping 
parameters. In the paramagnetic regime, which conserves time reversal 
symmetry, the model yields a three-fold degenerate GS, corresponding 
to the lowest-energy spin-multiplet $J=1$, describing the effective 
exchange antiferromagnetic coupling between the hole spin $j=3/2$ and 
the manganese spin $S=5/2$. The preservation of time reversal symmetry, 
which can be enforced in this many-body approach, is crucial. 
Note that on the general basis of group theory, irreducible representations 
of $S=1$ in a tetragonal symmetry must necessarily be degenerate. 
The case considered here is an example of this general property. 
This property is lost when the Mn quantum spin is replaced by a 
classical spin vector pointing in some arbitrary direction. Likewise 
and for the same reason, it is also lost in any mean-field treatment 
of the many-body Hamiltonian describing the Mn impurity. This is the 
case of our quantum Hamiltonian, where the $d-$levels are chosen to be 
spin-polarized in a similar arbitrary but fixed direction, effectively 
behaving like an external magnetic field, which breaks both time and 
rotational invariance. Note that this is also the case of \textit{ab-initio} 
spin density functional theory, whose rationale is quite close to our 
d-level model. The ensuing effective one-body Hamiltonian displays 
in three in-gap single-particle acceptor states of predominantly 
p-character, which are typically split by an amount of the order of 
the spin-orbit coupling strength. Thus the GS has degeneracy one 
instead of three, as derived in the many-body approach.\\
It turns out that this drawback of the effective one-particle 
Hamiltonian (either with the classical spin-model or with the 
mean-field treatment of the $d-$levels) and its incorrect description 
of the GS degeneracy is uninfluential when it comes to describing 
the properties of the Mn –acceptor state which can be probed by 
STM experiments, for example. The reason is that, when an electron 
is added to the system (via electron tunneling, for example), 
the inclusion of an interaction Hubbard U implies that only a 
single state is accessible even in the many-body approach. In 
other words, for typical values of the Hubbard U parameter, a 
second-bound state is not found for the impurity(see, for example, 
the discussion in Ref.~\onlinecite{tangflatte_prl04}). This important 
remark justifies the use of the effective single-particle models 
considered above, and explains their remarkable success in reproducing 
the main features of the Mn acceptor wave functions probed in STM experiments.\\
A perfect three-fold degenerate level is expected for the
present model and for the models of 
Ref.~\onlinecite{tangflatte_prl04, scm_MnGaAs_paper1_prb09} when 
SOI is switched off. 
Indeed, our calculations show that 
(i) the splitting between the three levels in the gap, as well as the 
relative position of the acceptor level with respect to the top 
of the valence band (113~meV) remain unchanged in very large clusters 
consisting of up to
30,000 atoms,
when SOI is included.
(ii) the small splitting still present in the supercell calculations  
with 3200 atoms {\it without} SOI is instead purely a finite-size effect, 
which vanishes when increasing the 
size of the supercell: the splitting reduces from 11.54~meV 
for 3200 atoms to 0.62~meV for 20,000 atoms. That is, in the absence of SOI,
the splitting is zero for this model.\\
Fig.~\ref{fig:MAE_BS} shows the magnetic-anisotropy-energy landscape for a single Mn in the bulk 
and on the (110) GaAs surface. The magnetic-anisotropy-energy landscape is defined as
the ground-state energy of the system plotted for different 
directions of the spin quantization axis, that is, as a function of the angles $\theta$ and $\phi$ 
(polar and azimuthal angles, which define the direction of the quantization axis). 
The coordinate system used for these plots has $\theta=0$ parallel
to the [001] direction and ($\theta=\frac{\pi}{2}$, $\phi=\frac{\pi}{2}$) parallel to [010] direction. 
We compare the results obtained with our TB model (top panels) and with the  
classical-spin model, introduced in Ref.~\onlinecite{scm_MnGaAs_paper1_prb09} (bottom panels). 
The magnetic anisotropy landscape 
is similar for the two models. In particular, for Mn in bulk the models feature 
two bistable easy axes, which are parallel to the
[001] direction and are separated by a barrier in the (001) plane 
({\it see} Figs.~\ref{fig:MAE_BS}(a) and (c)).
The fact that the shape of the magnetic anisotropy 
landscape does not change, means that the symmetry properties that 
control this quantity are correctly represented by the classical-spin 
model, or at least, they are captured in the same way as the more microscopic model.\\ 
Although the overall shapes of magnetic 
anisotropy landscapes for Mn on the surface are in qualitative agreement 
({\it see} Figs.~\ref{fig:MAE_BS}(b) and (d)),  
the anisotropy energy is one order of magnitude smaller in 
the case of our quantum $d$-level  
model. A smaller value of the anisotropy energy is consistent with the picture 
of a localized acceptor level (Fig.~\ref{fig:spec-LDOS}). 
The set of TB parameters extracted from GGA+U calculations 
for Mn on the surface gives a deeper and a more localized acceptor state,  
compared to the calculations done with the classical spin model~\cite{scm_MnGaAs_paper1_prb09}, which 
in turn leads to lower anisotropy.
\begin{figure}[htp]
\centering
\includegraphics[scale=0.22]{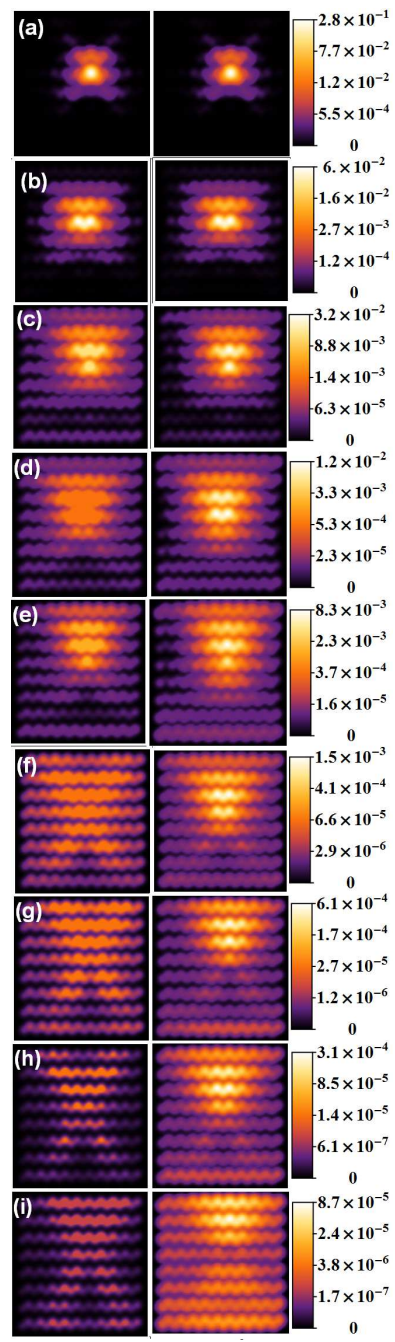}
\caption{Color online -- The (110) surface LDOS as a function of Mn depth. Panels (a) to (i) correspond to the LDOS 
 calculated for Mn in sublayers one to nine, respectively. The left and the right columns
show the LDOS for easy and hard direction, respectively.}
\label{fig:V-9LDOS}
\end{figure}
Figs.~\ref{fig:V-9LDOS} and \ref{fig:MAE-Layers} show the 
calculated LDOS and the anisotropy energy of the acceptor state when Mn is 
successively moved down from the surface layer towards the center of the cluster. 
Fig.~\ref{fig:MAE-Layers} shows a comparison of the anisotropy energy calculated with the two models as 
a function of the Mn depth, where we also include the values for the (110) surface and for the bulk.  
Here the value of the anisotropy energy is defined to be difference between 
the maximum and the minimum ground-state energy, 
calculated as a function of either the direction of magnetic moment in the 
case of classical-spin model, or the quantization axis in the quantum $d$-level model. 
We would like to point several important features 
that can be seen on this figure. The lower value of the anisotropy energy compared to the classical-spin 
model, found in our calculations, persists for almost all 
of the considered subsurfaces. The relatively high anisotropy for the bulk 
found in both models is a finite-size effect. In fact, our calculations demonstrate  
that the bulk anisotropy energy drops to a small fraction of a meV when the number of atoms 
in the cluster is increased by a factor of ten. 
As Mn is moved toward the center of the cluster, one expects the anisotropy 
energy to drop to its bulk value.
The present cluster (3200 atoms) includes 20 Ga layers along the [110] direction, therefore the ninth 
Ga sublayer is the last sublayer where we can replace a Ga with a Mn impurity atom. 
By increasing the size of the cluster we are  able to place the Mn atom in the deeper sublayers. 
As expected,  
such calculations show that the anisotropy energy 
decreases toward its bulk value as Mn is moved further away from the surface.\\
It is important to point out that the qualitative behavior of the magnetic 
anisotropy energy landscape (for the relevant case of a Mn close to the surface) 
remains unchanged with the size of the cluster. This includes the landscape of 
the anisotropy energy (e.g., easy and hard axes) as well as the anisotropy energy as a 
function of Mn depth (Fig.~\ref{fig:MAE-Layers} of the paper). Calculations carried out 
on much larger clusters~\cite{rm_JPCM} show that show 
that the qualitative behavior of magnetic anisotropy in 
Figs.~\ref{fig:MAE_BS} and~\ref{fig:MAE-Layers} remains intact as a function of the cluster size, while the value of 
anisotropy energy saturates to a smaller value without any qualitative change.\\
Note that the initial increase of the anisotropy energy up to the fifth sublayer, 
and its subsequent decrease, 
also reported in Ref.~\onlinecite{scm_MnGaAs_paper1_prb09}, is most likely due to the quasi-degeneracy between 
the last occupied and the acceptor states. It can be explained intuitively by looking at Fig.~\ref{fig:V-9LDOS}. 
The acceptor wave function becomes more extended as Mn moves away from the first sublayer. Such 
an extended wave function will be strongly affected by the surface until the Mn is moved 
deep enough so that the surface effects start to diminish (this corresponds to the sixth sublayer). 
A very small magnetic anisotropy of the first sublayer (of the order
of 0.06 meV) is due to its highly localized 
wave function ({\it see} Fig.~\ref{fig:V-9LDOS}(a)). 
The acceptor wave function in this case is less anisotropic, compared to 
the surface acceptor, which can be seen from the LDOS for 
easy and hard directions. The easy or hard direction here 
refer to a direction of the quantization axis for which the ground state energy of the system is minimum
or maximum, respectively.\\ 
The comparison between the anisotropy energy of Mn calculated with the two models seems to indicate that 
the difference is most pronounced on the surface and in the first five sublayers below the surface. 
As we mentioned earlier, this is due to the fact that the surface acceptor state in  
our fully quantum TB model is a deeper acceptor compared to its classical counterpart, 
which leads to a lower anisotropy energy.
\begin{figure}[htp]
\centering
\includegraphics[scale=0.25]{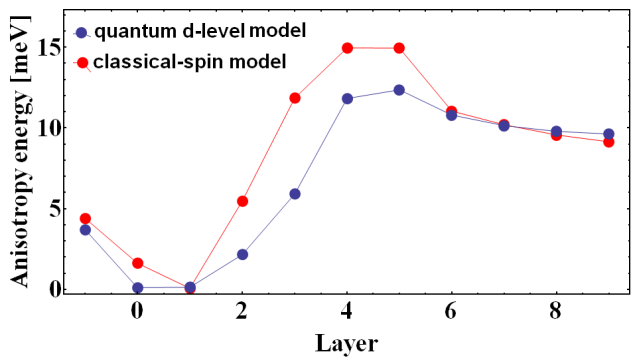}
\caption{Color online -- Magnetic anisotropy energy as 
a function of Mn depth. The values on the horizontal axis correspond to the sublayer index, in which the Mn impurity 
is located. The value at zero is the magnetic anisotropy energy of the (110) surface. The value   
before zero on the horizontal axis shows the magnetic anisotropy energy of the bulk. 
Red curve is for the case of the classical-spin model and blue curve shows the result 
obtained with the microscopic 
TB model, which includes the Mn $d$-levels (quantum $d$-level model).}
\label{fig:MAE-Layers}
\end{figure}
\begin{figure}[htp]
\centering
\includegraphics[scale=0.25]{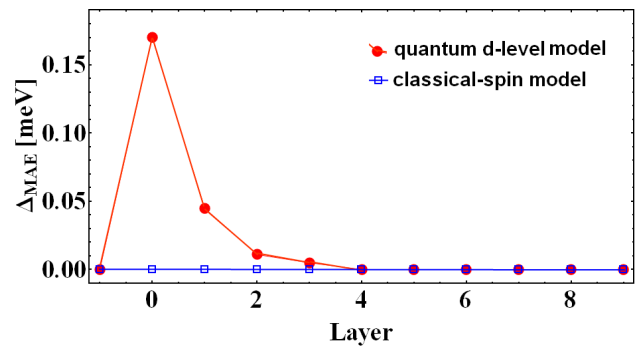}
\caption{Color online -- Difference $\Delta_{\rm  MAE}$ between the ground-state and the acceptor magnetic 
anisotropy energy as a function of Mn depth. See Eq.~\ref{Delta_mae2}
The notation on the horizontal axis is  the same as in Fig.~\ref{fig:MAE-Layers}.
The red line is the result for the quantum $d$-level model. The blue line ($\Delta_{\rm  MAE}= 0$) is 
the result for the classical-spin model.}
\label{fig:Mn-MAE-diff}
\end{figure}
Another difference between the two models is 
illustrated in Fig.~\ref{fig:Mn-MAE-diff}, 
where we plot the difference of the ground-state (GS) and 
the acceptor-level anisotropy energy, as a function of the Mn depth.
For the classical-spin model it was found that the 
energy of the (single-particle) acceptor level 
$\epsilon_{\rm acc}(\theta, \phi)$ and (many-particle) GS energy of the 
system $E(\theta, \phi)$ are very accurately related by
\begin{linenomath*}
\begin{equation}
\epsilon_{\rm acc}(\theta, \phi) 
= - E(\theta, \phi) + C\,,
\label{acc_anisotropy}
\end{equation}
\end{linenomath*}
where $C$ is a constant independent of $\theta$, $ \phi$.
This means that the sum of the two energies $E(\theta, \phi)$ and $\epsilon_{\rm acc}(\theta, \phi)$ is the same for  
any direction of the Mn magnetic moment. 
If $(\theta_{\rm max},  \phi_{\rm max})$  
and $(\theta_{\rm min},  \phi_{\rm min})$ define the two directions
where $E(\theta, \phi)$ attains its maximum and minimum value respectively,
from Eq.~\ref{acc_anisotropy} we obtain
\begin{linenomath*}
\begin{align}
&[E(\theta_{\rm max}, \phi_{\rm max})- E(\theta_{\rm min}, \phi_{\rm min})]  \nonumber\\
& +  [\epsilon_{\rm acc}(\theta_{\rm max}, \phi_{\rm max})- \epsilon_{\rm acc}(\theta_{\rm min}, \phi_{\rm min})] =0\;.
\label{Delta_mae}
\end{align}
\end{linenomath*}
The quantity $[E(\theta_{\rm max}, \phi_{\rm max})- E(\theta_{\rm min}, \phi_{\rm min})] $
is by definition the magnetic anisotropy of the system,  ${\rm MAE} $. Similarly,  Eq.~\ref{acc_anisotropy} implies that
$[\epsilon_{\rm acc}(\theta_{\rm max}, \phi_{\rm max})- \epsilon_{\rm acc}(\theta_{\rm min}, \phi_{\rm min})]$
is the {\it opposite} of the magnetic anisotropy of the acceptor level, $(-{\rm MAE})_{\rm acc}$.
Therefore, we can rewrite Eq.~\ref{Delta_mae} as
\begin{linenomath*}
\begin{equation}
\Delta_{\rm  MAE} \equiv {\rm MAE} - ({\rm MAE})_{\rm acc}= 0
\label{Delta_mae2}
\end{equation}
\end{linenomath*}
Eq.~\ref{acc_anisotropy} and Eq.~\ref{Delta_mae2} are very useful and powerful. 
They imply that the total anisotropy
of the system is essentially determined by the anisotropy of the single-particle acceptor level.
This picture remains valid as long as the coupling to the conduction band is not sensitive 
to the magnetization orientation.
It turns out that 
in the case of  
the quantum $d$-level TB model, 
Eq.~\ref{acc_anisotropy} is not exactly satisfied.
As a result, the quantity $\Delta_{\rm  MAE}$ is not exactly zero, 
although, as shown in Fig.~\ref{fig:Mn-MAE-diff}, its value is negligible 
for most of the cases, except for the surface and the first sublayers.\\ 
We suggest that the small change in the difference of the GS and the acceptor
anisotropy energy is due to the inclusion of the $d$-orbitals, which 
brings about a magnetization-direction dependence coupling with the conduction band.
In the classical-spin model, the majority $d$-electrons are essentially 
represented by a classical vector of fixed value $+5/2$ $\mu _B$, which only affects
the (occupied) energy-levels of the valence band through its SOI-induced orientation dependence.
In contrast, our quantum $d$-level model includes the impurity 
$d$-orbitals and the hopping between the $d$-orbitals and the nearest neighbor As atoms 
explicitly in the Hamiltonian. Unoccupied spin-down (minority) $d$-levels, 
located way up in the conduction
band, hybridize with like-spin As p-orbitals of the valence band. 
This coupling 
is responsible for the small deviation from Eq.~\ref{Delta_mae2},
which is also affected by the distance of the Mn atom from the surface.
as shown in Fig.~\ref{fig:Mn-MAE-diff}.
\subsection{Fe dopants on (110) GaAs surface}
\label{Fe dopants}
As we showed in section~\ref{Mn dopants}, despite quantitative differences between 
the classical and the fully quantum treatments of the impurity magnetic moment, 
the classical spin model gives a good estimate of magnetic and electronic properties 
of the  $[{\rm Mn}^{2+}, h]^0$ state, with a half-filled 3$d$-shell of Mn and a bound  
acceptor state. However, when the electronic transitions within the $d$-level shell of 
the dopant are important as in the case of Fe in GaAs, the inclusion of $3d$ electrons is necessary. 
In this section we present the results of our fully quantum microscopic TB model for 
Fe in the bulk and on the (110) surface of GaAs.
\begin{figure}[htp]
\centering
\includegraphics[scale=0.2]{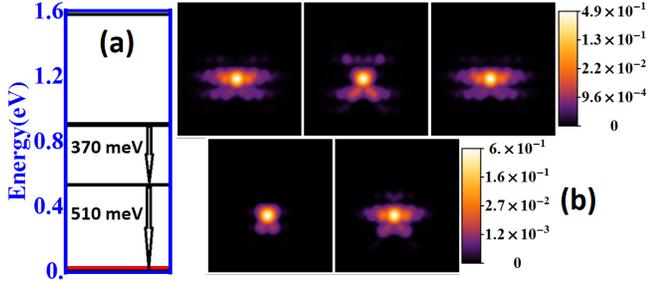}
\caption{Color online -- (a) Electronic structure of the Fe dopant in bulk 
GaAs showing the three $t_{2g}$ and the two $e_g$  
 levels inside the gap. The red line indicates the Fermi level. 
(b) The (110) cross-sectional LDOS for the five $d$-levels in the gap. Top panels are for 
the $t_{2g}$ triplet and the bottom panels are for the $e_g$ doublet.}
\label{fig:Figure3}
\end{figure}
Fe in the neutral acceptor state, $[{\rm Fe}^{2+}, h]^0$, in GaAs has six electrons in 
its $3d$ shell plus a weakly bound hole. The transition from $[{\rm Fe}^{2+}, h]^0$ 
to the neutral isoelectronic state $[{\rm Fe}^{3+}]^0$ occurs when the minority $d$-electron 
occupies the hole bound to Fe atom. A fully unoccupied minority $d$ orbitals in our TB 
represents the Fe atom in its isoelectronic state ($[{\rm Fe}^{3+}]^0$), 
while the neutral acceptor state 
($[{\rm Fe}^{2+}, h]^0$)  
is realized when one electron from the valance band occupies one of the minority $d$-orbitals, and 
creates an electron-hole excitation. ({\it see} Fig.~\ref{fig:elec}).\\
Fig.~\ref{fig:Figure3} shows the electronic structure and the LDOS of a single Fe 
impurity in the bulk. The minority doublet ($e_g$) level lies at 510~meV above  
the top of the valence band while the minority triplet ($t_{2g}$) level is found at 370~meV above 
the $e_g$ level. This bulk level-structure is in agreement with the results 
reported previously.~\cite{Malguth_phys_status, bocq_prb13}
The LDOS for each of the degenerate $e_g$ and $t_{2g}$ levels are shown in the bottom  
 and in the top of Fig.~\ref{fig:Figure3}(b), respectively. The wave functions of the 
  two-fold generate level are highly localized 
and 60\% of the spectral weight is located at the Fe site. In the case of the three-fold degenerate 
 level almost 50\% of the spectral weight is located on the Fe atom.
\begin{figure}[htp]
\centering
\includegraphics[scale=0.2]{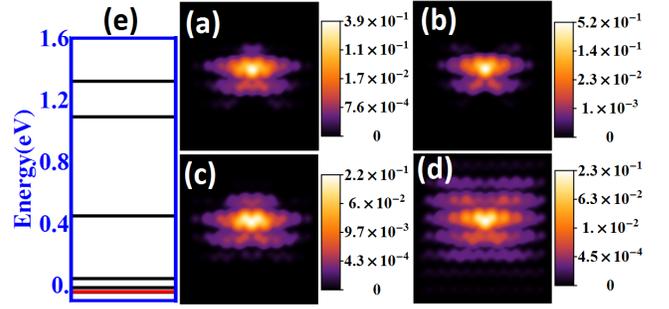}
\caption{Color online -- Electronic structure and LDOS for the Fe dopant on the (110) GaAs surface. 
(a-d) The (110) cross-sectional LDOS for the four top-most levels in the gap, with (a) being the 
 level with the highest energy and (d) the level with the lowest energy. (e) Electronic level-structure 
 showing the symmetry-broken $t_{2g}$ and $e_g$ levels inside the gap.  
The red line indicates the Fermi level.}
\label{fig:Fe-S}
\end{figure}
The lower symmetry of the surface compared to the bulk changes the electronic structure of the 
 impurity. Fig.~\ref{fig:Fe-S}(e) shows the energy level-structure inside the gap for Fe positioned  
on the (110) GaAs surface. As one can clearly see from the figure, 
all of the levels inside the gap, which are primarily originating from the Fe $d$-orbitals, are split. 
The two unoccupied levels immediately above the Fermi level 
 are closer to the valence band edge and are therefore more delocalized around 
the impurity. The splitting between these two levels is $\approx$60~meV. 
The LDOS calculated for this pair of states is plotted in Fig.~\ref{fig:Fe-S}(d). 
The other three levels in the gap appear at 0.45~eV, 1.1~eV and 1.3~eV. The corresponding LDOS images 
for each of the levels are plotted in Figs.~\ref{fig:Fe-S}(a-c). These distinct energy states 
appear to be more localized on the impurity compared to the states which are closer to the valence band edge.  
The shape of these acceptor states bear a striking resemblance to the STM topographies of Fe impurities on GaAs surface,    
  reported by Richardella \textit{et al.} ({\it see} Fig.~2(c) in Ref.~\onlinecite{yazdani_prb09}). However, we 
find differences in the exact positions of the levels in the gap, compared to the 
  experimental data. In fact, Richardella \textit{et al.} found two Fe-induced peaks 
	 in their spectroscopic data, located at $0.87\pm0.05$~eV and $1.52\pm0.05$~eV, respectively. 
	We note, however, that due to the finite energy resolution of the STM, the existence of more than 
one energy level within the width of a broad peak in the experimental spectroscopic data is not unlikely. 
In addition, the 
 positions of the peaks can be modified by the tip induced band bending. One should also take into account that the splitting between the 
 energy levels in the gap are sensitive to the amount of the lattice distortion.\\ 
The above considerations 
 may provide an explanation for the quantitative differences in the in-gap level-structure obtained with our TB model and that 
reported in Ref.~\onlinecite{yazdani_prb09}. Finally, we note that in a more recent experimental study~\cite{svenja_PRB_13}, the authors 
 found evidence of six peaks in their $dI/dV$ measurements for Fe on the (110) GaAs surface. In particular, one 
of the observed peaks can be related to the two closely spaced energy levels in the vicinity of the Fermi energy that 
we found in our TB calculations ({\it see} Fig.~\ref{fig:Fe-S}). The occupancy of the levels close to the Fermi energy 
can be different depending on the valence state of Fe, namely $[{\rm Fe}^{2+}, h]^0$ and $[{\rm Fe}^{3+}]^0$.\\ 
Fig.~\ref{fig:Fe-MAE_SB} shows the magnetic anisotropy energy  
of the Fe impurity in the $[{\rm Fe}^{3+}]^0$ state 
for different directions of the quantization axis. 
The surface anisotropy 
energy is approximately 1~meV, while Fe in 
bulk GaAs displays a considerably smaller anisotropy energy, 
of the order of $10^{-4}$~meV. 
The magnetic anisotropy energy landscape for the $[{\rm Fe}^{2+}, h]^0$ 
state (not shown here) is qualitatively and 
quantitatively similar to that of $[{\rm Mn}^{2+}, h]^0$.\\ 
In Fig.~\ref{fig:Fe-MAE_layer} 
we compare the anisotropy energy for the two valence states of Fe, $[{\rm Fe}^{3+}]^0$ and $[{\rm Fe}^{2+}, h]^0$, 
as the impurity atom is moved down from the (110) surface toward the
center of the cluster. 
The qualitative and quantitative difference between the 
anisotropy energy in the two valence states is remarkable.
While for  $[{\rm Fe}^{2+}, h]^0$  the sublayer dependence shows strong
similarities with the case of $[{\rm Mn}^{2+}, h]^0$, the
anisotropy for the the isoelectronic $[{\rm Fe}^{3+}]^0$ state is quite different: it is
typically
one order of magnitude smaller that for $[{\rm Fe}^{2+}, h]^0$ and depends 
weakly on the sublayer.
This striking behavior can be easily understood on the basis of the discussion
leading to Eqs.~\ref{acc_anisotropy}-\ref{Delta_mae2}. 
As for the Mn impurity, the total magnetic anisotropy
of the system is closely related to the magnetic anisotropy of the acceptor levels.
As seen in Fig.~\ref{fig:elec}, in the isoelectronic $[{\rm Fe}^{3+}]^0$ state 
this $p$-orbital-like acceptor level is occupied by the extra electron 
that Fe has with respect to Mn. 
Therefore the total energy of the system
contains also the contribution of this level. Then Eq.~\ref{Delta_mae2} 
implies that most of the
anisotropy coming from all the other occupied levels of the valance band is
essentially canceled by the approximately equal and opposite contribution coming from
the occupied acceptor. In contrast, in the $[{\rm Fe}^{2+}, h]^0$ valence state, the
extra electron occupies one of the higher minority $d$-orbital levels in the gap, and the 
acceptor level is empty (or is occupied by a "hole"). Thus, as for the $[{\rm Mn}^{2+}, h]^0$,
the cancellation does not occur. Furthermore, the anisotropy of the top-most $d$-orbital level
is, as expected, small. Therefore the behavior of $[{\rm Fe}^{2+}, h]^0$ is quite
similar to $[{\rm Mn}^{2+}, h]^0$. These simple considerations lead us to predict
that, at least within our non-self-consistent treatment, the charged 
states $[{\rm Fe}^{2+}]^{-}$ and $[{\rm Fe}^{3+}, h]^+$ should have the same behavior
of $[{\rm Fe}^{3+}]^0$ and $[{\rm Fe}^{2+}, h]^0$ respectively, since the
occupancy of the acceptor state is the same. This is exactly what our calculations
show. Again, we emphasize that our non-self consistent calculations should be
taken cautiously when charged states are involved.\\
The importance of this result stems from the fact that the valence and charge state
of individual TM impurities in GaAs can presently be manipulated with a variety of techniques.
For example, as mentioned earlier, it is possible to switch the impurity from 
the $[{\rm Fe}^{3+}]^0$ to the $[{\rm Fe}^{2+}]^{-}$ state 
via a voltage-dependent local manipulation
of the Fermi level 
by means of tip-induced band bending
in STM experiments.~\cite{bocq_prb13}
A similar manipulation of the Fermi level might soon allow the
switching between the $[{\rm Fe}^{3+}]^0$ and the $[{\rm Fe}^{3+}, h]^+$ charge state.
On the other hand, optical manipulation of the valence state might permit
switching between the $[{\rm Fe}^{3+}]^0$ and the $[{\rm Fe}^{2+}, h]^0$ states.
Our calculations indicate that the switching between these valence and charged
states should be also accompanied and characterized by significant changes in the
magnetic anisotropy energy of the system.\\
\begin{figure}[htp]
\centering
\includegraphics[scale=0.2]{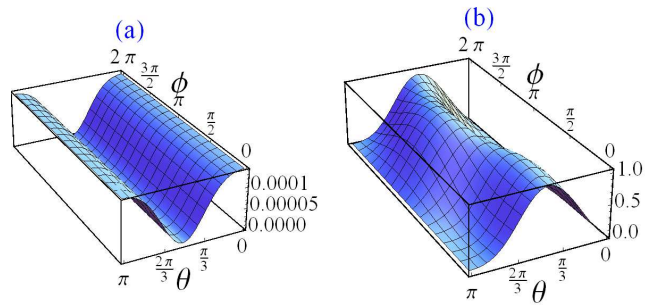}
\caption{Color online -- Magnetic anisotropy landscape for the Fe dopant in bulk GaAs (a) and Fe on the (110) GaAs surface (b). 
Energy is in the units of meV.}
\label{fig:Fe-MAE_SB}
\end{figure}
We would like to mention that for a Fe impurity in bulk, our \textit{ab-initio} calculations, 
in agreement with previously published results, show that the splitting 
of the (minority) $d-$levels in the GaAs gap follows the behavior described 
in Fig.~\ref{fig:elec}. This figure shows that the dominant effect of the splitting 
comes from the tetragonal symmetry of the lattice. Effects coming from 
the spin-orbit interaction, which are certainly present for the $d$ ($l=2$) 
minority-spin level occupied by the sixth electron in Fe, seem to be small. 
The situation for the Fe dopant on the surface is more complex since, as we have already 
described, in this case the surface-induced broken symmetry 
contributes substantially to the splitting, in a way that is not easily 
disentangled from other, more intrinsic, mechanisms. In any case, it is 
quite remarkable that the behavior of the anisotropy for the $[{\rm Fe}^{2+}, h]^0$ 
complex resembles qualitatively the behavior of the $[{\rm Mn}^{2+}, h]^0$: in both 
cases the total anisotropy is controlled  by the anisotropy of the impurity 
acceptor level, regardless of whether or not an extra electron is present.\\
\begin{figure}[htp]
\centering
\includegraphics[scale=0.25]{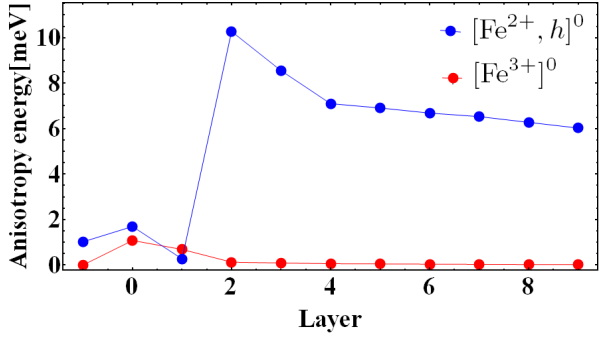}
\caption{Color online -- Magnetic anisotropy energy as a function of 
 the Fe depth. The notations on the horizontal are the same as in Fig.~\ref{fig:MAE-Layers}, 
with zero corresponding to the surface layer and the value below zero to the bulk result. Red dots are for the $[{\rm Fe}^{3+}]^{0}$ state 
and blue dots for the $[{\rm Fe}^{2+}, h]^0$ state.}
\label{fig:Fe-MAE_layer} 
\end{figure}
Fig.~\ref{fig:Fe-MAE} shows the magnetic anisotropy landscape 
 for the Fe impurity in the isoelectronic state as a function of its position below the surface. 
 Although the results for the surface layer (Fig.~\ref{fig:Fe-MAE_SB}(b)) and for the first sublayer 
(Fig.~\ref{fig:Fe-MAE_layer}(a)) are similar, in the latter case the magnetic anisotropy energy is smaller. 
As the Fe atom is moved away  
from the surface, the behavior of the magnetic anisotropy approaches that found in the bulk. 
When Fe is placed in the ninth sublayer, i.e in the middle of the cluster, 
 we expect it to behave like Fe in the bulk. This is confirmed in Fig.~\ref{fig:Fe-MAE}(i), which 
 shows that the 
 anisotropy energy has now decreased down to $10^{-2}$ meV compared to $1$~meV for the surface
 and the anisotropy landscape is almost identical to its bulk counterpart ({\it see} Fig.~\ref{fig:Fe-MAE_SB}(a)).
\begin{figure}[htp] 
\centering
\includegraphics[scale=0.2]{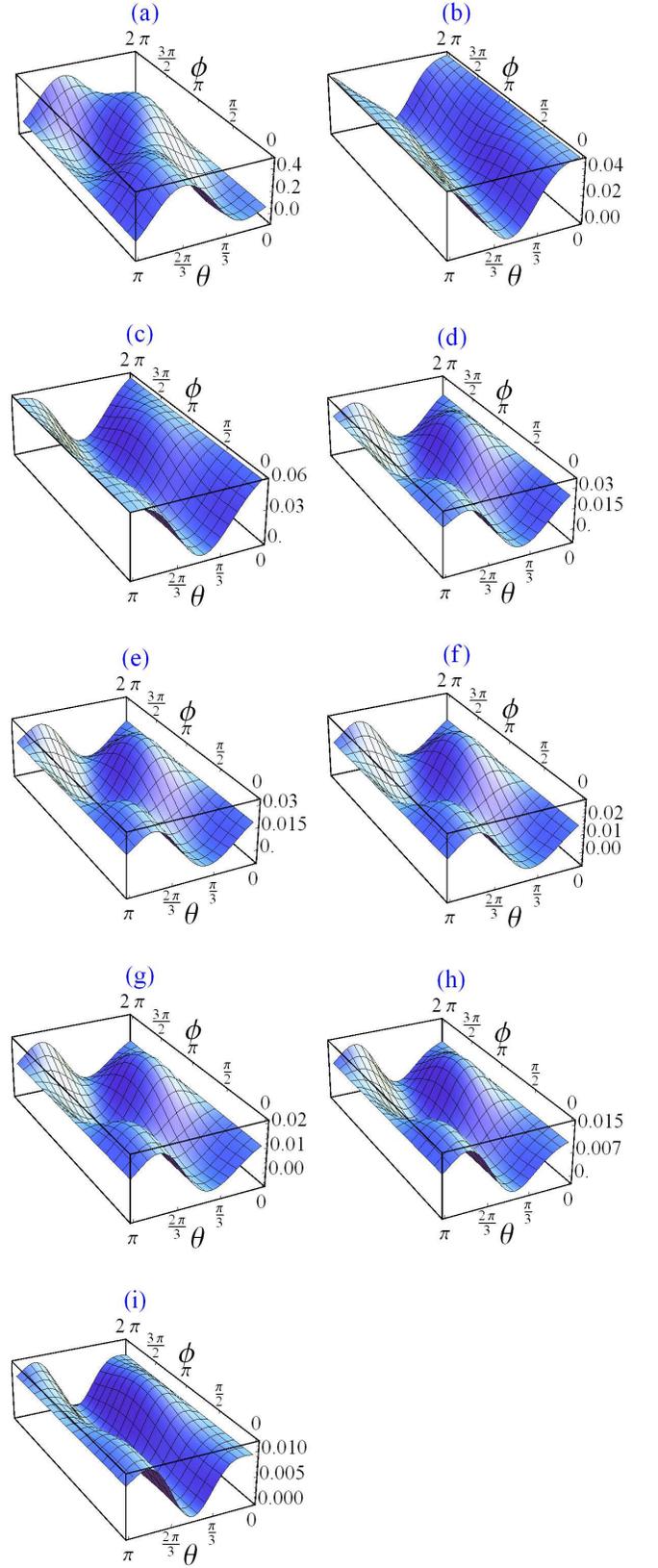}  
\caption{Color online -- Magnetic anisotropy energy landscapes for Fe in the $[Fe^{3+}]^0$ electronic configuration 
as a function of the impurity position below the surface. Panels (a) to (i) 
 correspond to Fe in the first to ninth sublayer, respectively. Energy is in the units of meV.}
\label{fig:Fe-MAE}
\end{figure}
Fig.~\ref{fig:LDOS-sublayers-Fe} shows the LDOS for the four 
top-most Fe-induced levels in the gap when the spin quantization axis is along the easy direction.  
As the Fe atom is moved down from the  
 surface towards the center of the cluster, the concentration of 
the spectral weight on the impurity site decreases and the LDOS becomes more delocalized. 
The butterfly shape of the LDOS around the impurity is more
pronounced for the sublayers located further away from the surface.
\begin{figure*}[htp]
\centering
\includegraphics[scale=0.25]{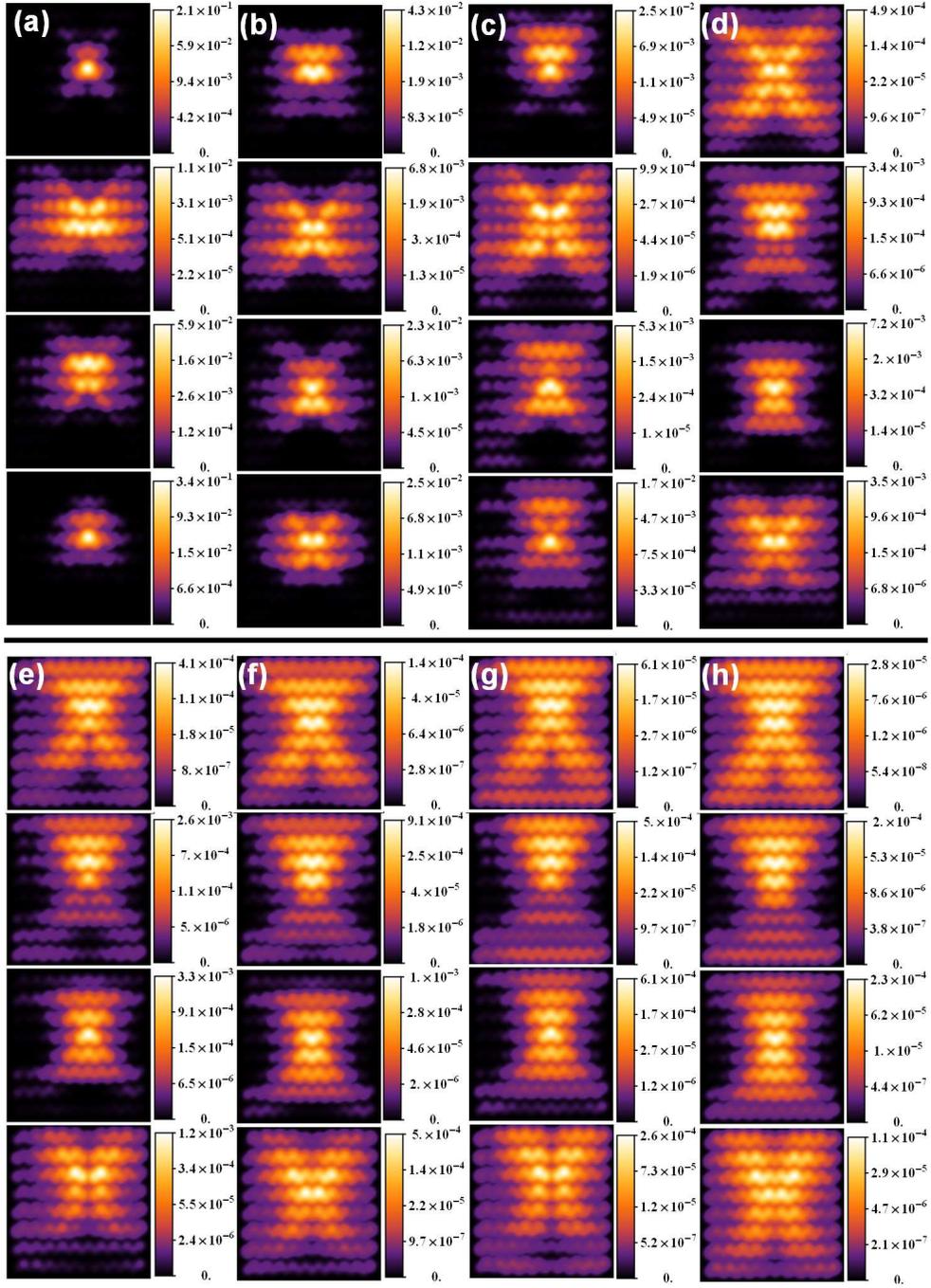}
\caption{Color online -- \label{fig:LDOS-sublayers-Fe}The (110) surface LDOS for four Fe-induced levels in the 
 gap as a function of the Fe depth. Panels (a) to (h) correspond to Fe in the first to eighth sublayer, respectively.}
\end{figure*}
The surface LDOS for Fe in the sublayers close to the surface is not very sensitive 
to the direction of the spin quantization axis. However, we   
find that when Fe is placed between the fifth and the eighth sublayer, the 
calculated LDOS images do display visible differences for the 
easy and hard directions, as shown in Fig.~\ref{fig:easy-hard}.  
These changes might well be detectable by STM, when the direction 
of the impurity magnetic moment is changed by an external magnetic field. 
An estimation based on the maximum anisotropy energy of 0.05~meV (for the 
fifth sublayer which decreases to 0.026~meV for the eighth 
sublayer) for the Fe on its isoelectronic state,  
suggests that it should be possible to manipulate the spin of the impurity with 
magnetic fields of the order of $10^{-1}$~T.  
This is very different from the case of Mn on the (110) GaAs surface. 
As was demonstrated in Ref.~\onlinecite{mc_MF_2013} both theoretically and experimentally, 
due to the strongly localized character of the Mn acceptor wavefunction on the surface 
and the large magnetic anisotropy energy of Mn in the near-surface layers, 
the acceptor hole LDOS is practically insensitive to the direction of the Mn magnetic moment 
in magnetic fields up to 6~T.  
In the case of Fe in its isoelectronic state, the combination of low anisotropy energy 
and the sensitivity of the surface LDOS  
to the direction of the impurity magnetic moment makes such manipulation possible with magnetic fields well within the 
experimental range.
\begin{figure}[htp]
\centering
\includegraphics[scale=0.25]{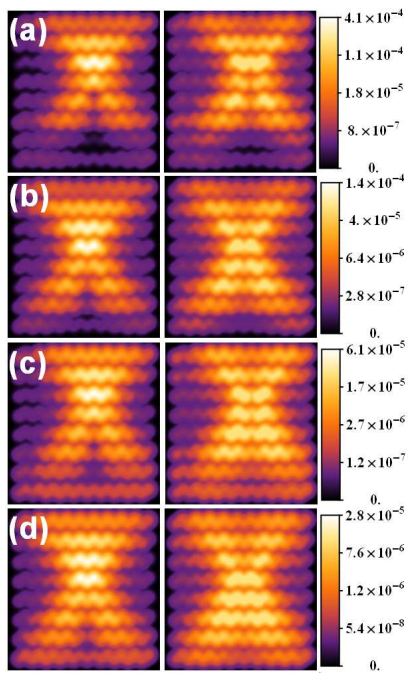}
\caption{Color online -- The (110) surface LDOS for the top-most level in the gap for two different direction 
 of the quantization axis. Panels (a) to (d) correspond to Fe in the fifth to eighth sublayer, respectively. 
The left(right) column shows the LDOS for the quantization axis along the easy(hard) direction.}
\label{fig:easy-hard}
\end{figure}
\section{Conclusions}
In this paper we have studied the electronic structure and 
magnetic properties of individual Mn and Fe dopants in the bulk  
and near the (110) surface of GaAs. Our theoretical treatment is 
based on a microscopic TB model including explicitly the 
$d$-orbitals of the dopant. 
We have employed DFT calculations to obtain the spin-resolved density 
of states for the impurity $d$-states, which was then used to 
determine the TB parameters for the  
$d$-orbitals. We calculated the in-gap electronic level-structure, LDOS 
and magnetic anisotropy landscapes for Mn and Fe impurities positioned 
on the surface or in subsurface layers.\\
Our calculations for Mn are typically in good agreement with the results 
obtained by a TB model where the impurity magnetic moment is treated
as an effective classical spin. In particular, we reproduced  
the well known features of Mn in GaAs, such as the position 
of the acceptor level in the gap and the spatial 
character of its wavefunction both in bulk and on the surface, which are also in agreement 
with experimental results. 
However, the microscopic quantum model finds a lower magnetic  
anisotropy energy for a Mn dopant on the surface compared to the classical-spin model. 
The difference between the anisotropy energy in the two models stems from the deeper
and therefore the considerably more localized character of the acceptor state on the surface, found 
in the quantum $d$-level model.\\
For the case of Fe in bulk, the microscopic model correctly finds
two degenerate minority-spin levels in the gap, of predominately $d$-character, 
with the expected two-fold and three-fold degeneracy associated to the $e_g$  and $t_{2g}$ 
symmetry respectively.
The structure of these minority $d$-character 
levels changes significantly when Fe is placed on the surface or in the nearby sublayers. 
Indeed, we find that the orbital degeneracy is lifted by surface effects,
and the electronic structure consists of
five unoccupied non-degenerate levels in the gap, two of which are very close to the top 
of valence band, two in the middle and one close to the conduction band. 
We were able to make connections between the calculated in-gap level structure 
and the experimental spectroscopic data for Fe on the (110) GaAs surface.\\
We also presented the calculated anisotropy energy landscapes and the LDOS of the in-gap states 
for Fe near the (110) surface. Importantly, 
we found that the anisotropy energy of the Fe impurity on the surface and 
subsurfaces depends on its valence state. 
Although the neutral acceptor state $[{\rm Fe}^{2+}, h]^0$ behaves similarly to Mn, for 
the isoelectronic state $[{\rm Fe}^{3+}]^0$ the anisotropy energy is considerably smaller, 
and behaves differently as a function of the impurity depth and the orientation of its 
magnetic moment. For a Fe dopant positioned in between the fifth to eighth sublayers, 
the anisotropy energy drops 
to a fraction of a meV. 
At the same time the spatial profile of the  
wavefunction associated with the top-most impurity level in the gap, is considerably
extended and therefore sensitive to the 
orientation of the impurity magnetic moment. 
This situation is quite different from the case of the Mn acceptor~\cite{mc_MF_2013}, 
whose wavefunction is strongly localized for a dopant on the surface (where the anisotropy is small)
and extended for a dopant in the sublayers (where, however, the anisotropy is large). 
These special
features of the Fe dopant will permit the manipulation of the Fe spin by means of an external magnetic 
field of few Tesla. Under this occurrence, the results of our calculations predict that, for a Fe dopant 
in its isoelectronic state, 
positioned a few monolayers 
below the (110) GaAs surface, a change in the direction of the dopant magnetic moment will produce 
a detectable difference in the STM cross-sectional view of the LDOS of the impurity  
levels in the gap.
\section*{Acknowledgments}
It is a pleasure to thank A.~H.~MacDonald, M.~E.~Flatt{\'e} and P.~M.~Koenraad 
for very useful discussions
and collaboration.
This work was
supported by the Faculty of Technology
at Linnaeus University, by the
Swedish Research Council under Grant Number: 621-2010-3761, 
and the NordForsk research network 080134 ``Nanospintronics: theory and
simulations".
Computational resources have
been provided by the Lunarc center for scientific and technical computing at
Lund University.
\bibliography{d_orbt}

\begin{thebibliography}{51}%
\makeatletter
\providecommand \@ifxundefined [1]{%
 \@ifx{#1\undefined}
}%
\providecommand \@ifnum [1]{%
 \ifnum #1\expandafter \@firstoftwo
 \else \expandafter \@secondoftwo
 \fi
}%
\providecommand \@ifx [1]{%
 \ifx #1\expandafter \@firstoftwo
 \else \expandafter \@secondoftwo
 \fi
}%
\providecommand \natexlab [1]{#1}%
\providecommand \enquote  [1]{``#1''}%
\providecommand \bibnamefont  [1]{#1}%
\providecommand \bibfnamefont [1]{#1}%
\providecommand \citenamefont [1]{#1}%
\providecommand \href@noop [0]{\@secondoftwo}%
\providecommand \href [0]{\begingroup \@sanitize@url \@href}%
\providecommand \@href[1]{\@@startlink{#1}\@@href}%
\providecommand \@@href[1]{\endgroup#1\@@endlink}%
\providecommand \@sanitize@url [0]{\catcode `\\12\catcode `\$12\catcode
  `\&12\catcode `\#12\catcode `\^12\catcode `\_12\catcode `\%12\relax}%
\providecommand \@@startlink[1]{}%
\providecommand \@@endlink[0]{}%
\providecommand \url  [0]{\begingroup\@sanitize@url \@url }%
\providecommand \@url [1]{\endgroup\@href {#1}{\urlprefix }}%
\providecommand \urlprefix  [0]{URL }%
\providecommand \Eprint [0]{\href }%
\providecommand \doibase [0]{http://dx.doi.org/}%
\providecommand \selectlanguage [0]{\@gobble}%
\providecommand \bibinfo  [0]{\@secondoftwo}%
\providecommand \bibfield  [0]{\@secondoftwo}%
\providecommand \translation [1]{[#1]}%
\providecommand \BibitemOpen [0]{}%
\providecommand \bibitemStop [0]{}%
\providecommand \bibitemNoStop [0]{.\EOS\space}%
\providecommand \EOS [0]{\spacefactor3000\relax}%
\providecommand \BibitemShut  [1]{\csname bibitem#1\endcsname}%
\let\auto@bib@innerbib\@empty
\bibitem [{\citenamefont {Koenraad}\ and\ \citenamefont
  {Flatt{\'e}}(2011)}]{pm_nam11}%
  \BibitemOpen
  \bibfield  {author} {\bibinfo {author} {\bibfnamefont {P.~M.}\ \bibnamefont
  {Koenraad}}\ and\ \bibinfo {author} {\bibfnamefont {M.~E.}\ \bibnamefont
  {Flatt{\'e}}},\ }\href@noop {} {\bibfield  {journal} {\bibinfo  {journal}
  {Nat. Mater.}\ }\textbf {\bibinfo {volume} {10}},\ \bibinfo {pages} {91}
  (\bibinfo {year} {2011})}\BibitemShut {NoStop}%
\bibitem [{\citenamefont {Yakunin}\ \emph {et~al.}(2004)\citenamefont
  {Yakunin}, \citenamefont {Silov}, \citenamefont {Koenraad}, \citenamefont
  {Wolter}, \citenamefont {Van~Roy}, \citenamefont {De~Boeck}, \citenamefont
  {Tang},\ and\ \citenamefont {Flatt\'e}}]{yakunin_prl04}%
  \BibitemOpen
  \bibfield  {author} {\bibinfo {author} {\bibfnamefont {A.~M.}\ \bibnamefont
  {Yakunin}}, \bibinfo {author} {\bibfnamefont {A.~Y.}\ \bibnamefont {Silov}},
  \bibinfo {author} {\bibfnamefont {P.~M.}\ \bibnamefont {Koenraad}}, \bibinfo
  {author} {\bibfnamefont {J.~H.}\ \bibnamefont {Wolter}}, \bibinfo {author}
  {\bibfnamefont {W.}~\bibnamefont {Van~Roy}}, \bibinfo {author} {\bibfnamefont
  {J.}~\bibnamefont {De~Boeck}}, \bibinfo {author} {\bibfnamefont {J.-M.}\
  \bibnamefont {Tang}}, \ and\ \bibinfo {author} {\bibfnamefont {M.~E.}\
  \bibnamefont {Flatt\'e}},\ }\href {\doibase 10.1103/PhysRevLett.92.216806}
  {\bibfield  {journal} {\bibinfo  {journal} {Phys. Rev. Lett.}\ }\textbf
  {\bibinfo {volume} {92}},\ \bibinfo {pages} {216806} (\bibinfo {year}
  {2004})}\BibitemShut {NoStop}%
\bibitem [{\citenamefont {Shinada}\ \emph {et~al.}(2005)\citenamefont
  {Shinada}, \citenamefont {Okamoto}, \citenamefont {Kobayashi},\ and\
  \citenamefont {Ohdomari}}]{shinada_nat05}%
  \BibitemOpen
  \bibfield  {author} {\bibinfo {author} {\bibfnamefont {T.}~\bibnamefont
  {Shinada}}, \bibinfo {author} {\bibfnamefont {S.}~\bibnamefont {Okamoto}},
  \bibinfo {author} {\bibfnamefont {T.}~\bibnamefont {Kobayashi}}, \ and\
  \bibinfo {author} {\bibfnamefont {I.}~\bibnamefont {Ohdomari}},\ }\href@noop
  {} {\bibfield  {journal} {\bibinfo  {journal} {Nature (London)}\ }\textbf
  {\bibinfo {volume} {437}},\ \bibinfo {pages} {1128} (\bibinfo {year}
  {2005})}\BibitemShut {NoStop}%
\bibitem [{\citenamefont {Kitchen}\ \emph {et~al.}(2006)\citenamefont
  {Kitchen}, \citenamefont {Richardella}, \citenamefont {Tang}, \citenamefont
  {Flatt{\'e}},\ and\ \citenamefont {Yazdani}}]{yazdani_nat06}%
  \BibitemOpen
  \bibfield  {author} {\bibinfo {author} {\bibfnamefont {D.}~\bibnamefont
  {Kitchen}}, \bibinfo {author} {\bibfnamefont {A.}~\bibnamefont
  {Richardella}}, \bibinfo {author} {\bibfnamefont {J.-M.}\ \bibnamefont
  {Tang}}, \bibinfo {author} {\bibfnamefont {M.~E.}\ \bibnamefont
  {Flatt{\'e}}}, \ and\ \bibinfo {author} {\bibfnamefont {A.}~\bibnamefont
  {Yazdani}},\ }\href@noop {} {\bibfield  {journal} {\bibinfo  {journal}
  {Nature}\ }\textbf {\bibinfo {volume} {442}},\ \bibinfo {pages} {436}
  (\bibinfo {year} {2006})}\BibitemShut {NoStop}%
\bibitem [{\citenamefont {Marczinowski}\ \emph {et~al.}(2007)\citenamefont
  {Marczinowski}, \citenamefont {Wiebe}, \citenamefont {Tang}, \citenamefont
  {Flatt\'e}, \citenamefont {Meier}, \citenamefont {Morgenstern},\ and\
  \citenamefont {Wiesendanger}}]{wiesendanger_MnInAs}%
  \BibitemOpen
  \bibfield  {author} {\bibinfo {author} {\bibfnamefont {F.}~\bibnamefont
  {Marczinowski}}, \bibinfo {author} {\bibfnamefont {J.}~\bibnamefont {Wiebe}},
  \bibinfo {author} {\bibfnamefont {J.-M.}\ \bibnamefont {Tang}}, \bibinfo
  {author} {\bibfnamefont {M.~E.}\ \bibnamefont {Flatt\'e}}, \bibinfo {author}
  {\bibfnamefont {F.}~\bibnamefont {Meier}}, \bibinfo {author} {\bibfnamefont
  {M.}~\bibnamefont {Morgenstern}}, \ and\ \bibinfo {author} {\bibfnamefont
  {R.}~\bibnamefont {Wiesendanger}},\ }\href {\doibase
  10.1103/PhysRevLett.99.157202} {\bibfield  {journal} {\bibinfo  {journal}
  {Phys. Rev. Lett.}\ }\textbf {\bibinfo {volume} {99}},\ \bibinfo {pages}
  {157202} (\bibinfo {year} {2007})}\BibitemShut {NoStop}%
\bibitem [{\citenamefont {Lee}\ and\ \citenamefont
  {Gupta}(2010)}]{gupta_science_2010}%
  \BibitemOpen
  \bibfield  {author} {\bibinfo {author} {\bibfnamefont {D.~H.}\ \bibnamefont
  {Lee}}\ and\ \bibinfo {author} {\bibfnamefont {J.~A.}\ \bibnamefont
  {Gupta}},\ }\href@noop {} {\bibfield  {journal} {\bibinfo  {journal}
  {Science}\ }\textbf {\bibinfo {volume} {330}},\ \bibinfo {pages} {1807}
  (\bibinfo {year} {2010})}\BibitemShut {NoStop}%
\bibitem [{\citenamefont {Garleff}\ \emph {et~al.}(2010)\citenamefont
  {Garleff}, \citenamefont {Wijnheijmer}, \citenamefont {Silov}, \citenamefont
  {van Bree}, \citenamefont {Van~Roy}, \citenamefont {Tang}, \citenamefont
  {Flatt\'e},\ and\ \citenamefont {Koenraad}}]{garleff_prb_2010}%
  \BibitemOpen
  \bibfield  {author} {\bibinfo {author} {\bibfnamefont {J.~K.}\ \bibnamefont
  {Garleff}}, \bibinfo {author} {\bibfnamefont {A.~P.}\ \bibnamefont
  {Wijnheijmer}}, \bibinfo {author} {\bibfnamefont {A.~Y.}\ \bibnamefont
  {Silov}}, \bibinfo {author} {\bibfnamefont {J.}~\bibnamefont {van Bree}},
  \bibinfo {author} {\bibfnamefont {W.}~\bibnamefont {Van~Roy}}, \bibinfo
  {author} {\bibfnamefont {J.-M.}\ \bibnamefont {Tang}}, \bibinfo {author}
  {\bibfnamefont {M.~E.}\ \bibnamefont {Flatt\'e}}, \ and\ \bibinfo {author}
  {\bibfnamefont {P.~M.}\ \bibnamefont {Koenraad}},\ }\href {\doibase
  10.1103/PhysRevB.82.035303} {\bibfield  {journal} {\bibinfo  {journal} {Phys.
  Rev. B}\ }\textbf {\bibinfo {volume} {82}},\ \bibinfo {pages} {035303}
  (\bibinfo {year} {2010})}\BibitemShut {NoStop}%
\bibitem [{\citenamefont {Fuechsle}\ \emph {et~al.}(2012)\citenamefont
  {Fuechsle}, \citenamefont {Miwa}, \citenamefont {Mahapatra}, \citenamefont
  {Ryu}, \citenamefont {Lee}, \citenamefont {Warschkow}, \citenamefont
  {Hollenberg}, \citenamefont {Klimeck},\ and\ \citenamefont
  {Simmons}}]{Fuec_nnt12}%
  \BibitemOpen
  \bibfield  {author} {\bibinfo {author} {\bibfnamefont {M.}~\bibnamefont
  {Fuechsle}}, \bibinfo {author} {\bibfnamefont {J.~A.}\ \bibnamefont {Miwa}},
  \bibinfo {author} {\bibfnamefont {S.}~\bibnamefont {Mahapatra}}, \bibinfo
  {author} {\bibfnamefont {H.}~\bibnamefont {Ryu}}, \bibinfo {author}
  {\bibfnamefont {S.}~\bibnamefont {Lee}}, \bibinfo {author} {\bibfnamefont
  {O.}~\bibnamefont {Warschkow}}, \bibinfo {author} {\bibfnamefont {L.~C.~L.}\
  \bibnamefont {Hollenberg}}, \bibinfo {author} {\bibfnamefont
  {G.}~\bibnamefont {Klimeck}}, \ and\ \bibinfo {author} {\bibfnamefont
  {M.~Y.}\ \bibnamefont {Simmons}},\ }\href@noop {} {\bibfield  {journal}
  {\bibinfo  {journal} {Nat. Nanotechnol.}\ }\textbf {\bibinfo {volume} {7}},\
  \bibinfo {pages} {242} (\bibinfo {year} {2012})}\BibitemShut {NoStop}%
\bibitem [{\citenamefont {Pla}\ \emph {et~al.}(2012)\citenamefont {Pla},
  \citenamefont {Tan}, \citenamefont {Dehollain}, \citenamefont {Lim},
  \citenamefont {Morton}, \citenamefont {Jamieson}, \citenamefont {Dzurak},\
  and\ \citenamefont {Morello}}]{pla_nat12}%
  \BibitemOpen
  \bibfield  {author} {\bibinfo {author} {\bibfnamefont {J.~J.}\ \bibnamefont
  {Pla}}, \bibinfo {author} {\bibfnamefont {K.~Y.}\ \bibnamefont {Tan}},
  \bibinfo {author} {\bibfnamefont {J.~P.}\ \bibnamefont {Dehollain}}, \bibinfo
  {author} {\bibfnamefont {W.~H.}\ \bibnamefont {Lim}}, \bibinfo {author}
  {\bibfnamefont {J.}~\bibnamefont {Morton}}, \bibinfo {author} {\bibfnamefont
  {D.~N.}\ \bibnamefont {Jamieson}}, \bibinfo {author} {\bibfnamefont {A.~S.}\
  \bibnamefont {Dzurak}}, \ and\ \bibinfo {author} {\bibfnamefont
  {A.}~\bibnamefont {Morello}},\ }\href@noop {} {\bibfield  {journal} {\bibinfo
   {journal} {Nature (London)}\ }\textbf {\bibinfo {volume} {489}},\ \bibinfo
  {pages} {541} (\bibinfo {year} {2012})}\BibitemShut {NoStop}%
\bibitem [{\citenamefont {Zhao}\ \emph {et~al.}(2004)\citenamefont {Zhao},
  \citenamefont {Mahadevan},\ and\ \citenamefont {Zunger}}]{zhao_apl04}%
  \BibitemOpen
  \bibfield  {author} {\bibinfo {author} {\bibfnamefont {Y.}~\bibnamefont
  {Zhao}}, \bibinfo {author} {\bibfnamefont {P.}~\bibnamefont {Mahadevan}}, \
  and\ \bibinfo {author} {\bibfnamefont {A.}~\bibnamefont {Zunger}},\
  }\href@noop {} {\bibfield  {journal} {\bibinfo  {journal} {Apl. Phys. Lett.}\
  }\textbf {\bibinfo {volume} {19}},\ \bibinfo {pages} {3753} (\bibinfo {year}
  {2004})}\BibitemShut {NoStop}%
\bibitem [{\citenamefont {Mahadevan}\ \emph {et~al.}(2004)\citenamefont
  {Mahadevan}, \citenamefont {Zunger},\ and\ \citenamefont
  {Sarma}}]{sarma_prl04}%
  \BibitemOpen
  \bibfield  {author} {\bibinfo {author} {\bibfnamefont {P.}~\bibnamefont
  {Mahadevan}}, \bibinfo {author} {\bibfnamefont {A.}~\bibnamefont {Zunger}}, \
  and\ \bibinfo {author} {\bibfnamefont {D.~D.}\ \bibnamefont {Sarma}},\
  }\href@noop {} {\bibfield  {journal} {\bibinfo  {journal} {Phys. Rev. Lett.}\
  }\textbf {\bibinfo {volume} {93}},\ \bibinfo {pages} {177201} (\bibinfo
  {year} {2004})}\BibitemShut {NoStop}%
\bibitem [{\citenamefont {Mikkelsen}\ \emph {et~al.}(2004)\citenamefont
  {Mikkelsen}, \citenamefont {Sanyal}, \citenamefont {Sadowski}, \citenamefont
  {Ouattara}, \citenamefont {Kanski}, \citenamefont {Mirbt}, \citenamefont
  {Eriksson},\ and\ \citenamefont {Lundgren}}]{PhysRevB.70.085411}%
  \BibitemOpen
  \bibfield  {author} {\bibinfo {author} {\bibfnamefont {A.}~\bibnamefont
  {Mikkelsen}}, \bibinfo {author} {\bibfnamefont {B.}~\bibnamefont {Sanyal}},
  \bibinfo {author} {\bibfnamefont {J.}~\bibnamefont {Sadowski}}, \bibinfo
  {author} {\bibfnamefont {L.}~\bibnamefont {Ouattara}}, \bibinfo {author}
  {\bibfnamefont {J.}~\bibnamefont {Kanski}}, \bibinfo {author} {\bibfnamefont
  {S.}~\bibnamefont {Mirbt}}, \bibinfo {author} {\bibfnamefont
  {O.}~\bibnamefont {Eriksson}}, \ and\ \bibinfo {author} {\bibfnamefont
  {E.}~\bibnamefont {Lundgren}},\ }\href {\doibase 10.1103/PhysRevB.70.085411}
  {\bibfield  {journal} {\bibinfo  {journal} {Phys. Rev. B}\ }\textbf {\bibinfo
  {volume} {70}},\ \bibinfo {pages} {085411} (\bibinfo {year}
  {2004})}\BibitemShut {NoStop}%
\bibitem [{\citenamefont {Stroppa}\ \emph {et~al.}(2007)\citenamefont
  {Stroppa}, \citenamefont {Duan}, \citenamefont {Peressi}, \citenamefont
  {Furlanetto},\ and\ \citenamefont {Modesti}}]{PhysRevB.75.195335}%
  \BibitemOpen
  \bibfield  {author} {\bibinfo {author} {\bibfnamefont {A.}~\bibnamefont
  {Stroppa}}, \bibinfo {author} {\bibfnamefont {X.}~\bibnamefont {Duan}},
  \bibinfo {author} {\bibfnamefont {M.}~\bibnamefont {Peressi}}, \bibinfo
  {author} {\bibfnamefont {D.}~\bibnamefont {Furlanetto}}, \ and\ \bibinfo
  {author} {\bibfnamefont {S.}~\bibnamefont {Modesti}},\ }\href {\doibase
  10.1103/PhysRevB.75.195335} {\bibfield  {journal} {\bibinfo  {journal} {Phys.
  Rev. B}\ }\textbf {\bibinfo {volume} {75}},\ \bibinfo {pages} {195335}
  (\bibinfo {year} {2007})}\BibitemShut {NoStop}%
\bibitem [{\citenamefont {Ebert}\ and\ \citenamefont
  {Mankovsky}(2009)}]{ebert_prb09}%
  \BibitemOpen
  \bibfield  {author} {\bibinfo {author} {\bibfnamefont {H.}~\bibnamefont
  {Ebert}}\ and\ \bibinfo {author} {\bibfnamefont {S.}~\bibnamefont
  {Mankovsky}},\ }\href@noop {} {\bibfield  {journal} {\bibinfo  {journal}
  {Phys. Rev. B}\ }\textbf {\bibinfo {volume} {79}},\ \bibinfo {pages} {045209}
  (\bibinfo {year} {2009})}\BibitemShut {NoStop}%
\bibitem [{\citenamefont {Islam}\ and\ \citenamefont
  {Canali}(2012)}]{fi_cmc_prb_2012}%
  \BibitemOpen
  \bibfield  {author} {\bibinfo {author} {\bibfnamefont {M.~F.}\ \bibnamefont
  {Islam}}\ and\ \bibinfo {author} {\bibfnamefont {C.~M.}\ \bibnamefont
  {Canali}},\ }\href@noop {} {\bibfield  {journal} {\bibinfo  {journal} {Phys.
  Rev. B}\ }\textbf {\bibinfo {volume} {85}},\ \bibinfo {pages} {155306}
  (\bibinfo {year} {2012})}\BibitemShut {NoStop}%
\bibitem [{\citenamefont {Tang}\ and\ \citenamefont
  {Flatt{\'e}}(2004)}]{tangflatte_prl04}%
  \BibitemOpen
  \bibfield  {author} {\bibinfo {author} {\bibfnamefont {J.-M.}\ \bibnamefont
  {Tang}}\ and\ \bibinfo {author} {\bibfnamefont {M.~E.}\ \bibnamefont
  {Flatt{\'e}}},\ }\href@noop {} {\bibfield  {journal} {\bibinfo  {journal}
  {Phys. Rev. Lett.}\ }\textbf {\bibinfo {volume} {92}},\ \bibinfo {pages}
  {047201} (\bibinfo {year} {2004})}\BibitemShut {NoStop}%
\bibitem [{\citenamefont {Tang}\ and\ \citenamefont
  {Flatt{\'e}}(2005)}]{tangflatte_prb05}%
  \BibitemOpen
  \bibfield  {author} {\bibinfo {author} {\bibfnamefont {J.-M.}\ \bibnamefont
  {Tang}}\ and\ \bibinfo {author} {\bibfnamefont {M.~E.}\ \bibnamefont
  {Flatt{\'e}}},\ }\href@noop {} {\bibfield  {journal} {\bibinfo  {journal}
  {Phys. Rev. B}\ }\textbf {\bibinfo {volume} {72}},\ \bibinfo {pages} {161315}
  (\bibinfo {year} {2005})}\BibitemShut {NoStop}%
\bibitem [{\citenamefont {Timm}\ and\ \citenamefont
  {MacDonald}(2005)}]{timmacd_prb05}%
  \BibitemOpen
  \bibfield  {author} {\bibinfo {author} {\bibfnamefont {C.}~\bibnamefont
  {Timm}}\ and\ \bibinfo {author} {\bibfnamefont {A.~H.}\ \bibnamefont
  {MacDonald}},\ }\href@noop {} {\bibfield  {journal} {\bibinfo  {journal}
  {Phys. Rev. B}\ }\textbf {\bibinfo {volume} {71}},\ \bibinfo {pages} {155206}
  (\bibinfo {year} {2005})}\BibitemShut {NoStop}%
\bibitem [{\citenamefont {Jancu}\ \emph {et~al.}(2008)\citenamefont {Jancu},
  \citenamefont {Girard}, \citenamefont {Nestoklon}, \citenamefont {Lemaitre},
  \citenamefont {Glas}, \citenamefont {Wang},\ and\ \citenamefont
  {Voisin}}]{Jancu_PRL_08}%
  \BibitemOpen
  \bibfield  {author} {\bibinfo {author} {\bibfnamefont {J.-M.}\ \bibnamefont
  {Jancu}}, \bibinfo {author} {\bibfnamefont {J.-C.}\ \bibnamefont {Girard}},
  \bibinfo {author} {\bibfnamefont {M.~O.}\ \bibnamefont {Nestoklon}}, \bibinfo
  {author} {\bibfnamefont {A.}~\bibnamefont {Lemaitre}}, \bibinfo {author}
  {\bibfnamefont {F.}~\bibnamefont {Glas}}, \bibinfo {author} {\bibfnamefont
  {Z.~Z.}\ \bibnamefont {Wang}}, \ and\ \bibinfo {author} {\bibfnamefont
  {P.}~\bibnamefont {Voisin}},\ }\href@noop {} {\bibfield  {journal} {\bibinfo
  {journal} {Physical Review Letters}\ }\textbf {\bibinfo {volume} {101}},\
  \bibinfo {pages} {196801} (\bibinfo {year} {2008})}\BibitemShut {NoStop}%
\bibitem [{\citenamefont {Strandberg}\ \emph {et~al.}(2009)\citenamefont
  {Strandberg}, \citenamefont {Canali},\ and\ \citenamefont
  {MacDonald}}]{scm_MnGaAs_paper1_prb09}%
  \BibitemOpen
  \bibfield  {author} {\bibinfo {author} {\bibfnamefont {T.~O.}\ \bibnamefont
  {Strandberg}}, \bibinfo {author} {\bibfnamefont {C.~M.}\ \bibnamefont
  {Canali}}, \ and\ \bibinfo {author} {\bibfnamefont {A.~H.}\ \bibnamefont
  {MacDonald}},\ }\href@noop {} {\bibfield  {journal} {\bibinfo  {journal}
  {Phys. Rev. B}\ }\textbf {\bibinfo {volume} {80}},\ \bibinfo {pages} {024425}
  (\bibinfo {year} {2009})}\BibitemShut {NoStop}%
\bibitem [{\citenamefont {Strandberg}\ \emph {et~al.}(2010)\citenamefont
  {Strandberg}, \citenamefont {Canali},\ and\ \citenamefont
  {MacDonald}}]{scm_MnGaAs_paper2_prb2010}%
  \BibitemOpen
  \bibfield  {author} {\bibinfo {author} {\bibfnamefont {T.~O.}\ \bibnamefont
  {Strandberg}}, \bibinfo {author} {\bibfnamefont {C.~M.}\ \bibnamefont
  {Canali}}, \ and\ \bibinfo {author} {\bibfnamefont {A.~H.}\ \bibnamefont
  {MacDonald}},\ }\href@noop {} {\bibfield  {journal} {\bibinfo  {journal}
  {Phys. Rev. B}\ }\textbf {\bibinfo {volume} {81}},\ \bibinfo {pages} {054401}
  (\bibinfo {year} {2010})}\BibitemShut {NoStop}%
\bibitem [{\citenamefont {Strandberg}\ \emph {et~al.}(2011)\citenamefont
  {Strandberg}, \citenamefont {Canali},\ and\ \citenamefont
  {MacDonald}}]{scm_MnGaAs_paper3_prl011}%
  \BibitemOpen
  \bibfield  {author} {\bibinfo {author} {\bibfnamefont {T.~O.}\ \bibnamefont
  {Strandberg}}, \bibinfo {author} {\bibfnamefont {C.~M.}\ \bibnamefont
  {Canali}}, \ and\ \bibinfo {author} {\bibfnamefont {A.~H.}\ \bibnamefont
  {MacDonald}},\ }\href@noop {} {\bibfield  {journal} {\bibinfo  {journal}
  {Phys. Rev. Lett.}\ }\textbf {\bibinfo {volume} {106}},\ \bibinfo {pages}
  {017202} (\bibinfo {year} {2011})}\BibitemShut {NoStop}%
\bibitem [{\citenamefont {Ma\ifmmode~\check{s}\else \v{s}\fi{}ek}\ \emph
  {et~al.}(2010)\citenamefont {Ma\ifmmode~\check{s}\else \v{s}\fi{}ek},
  \citenamefont {M\'aca}, \citenamefont {Kudrnovsk\'y}, \citenamefont
  {Makarovsky}, \citenamefont {Eaves}, \citenamefont {Campion}, \citenamefont
  {Edmonds}, \citenamefont {Rushforth}, \citenamefont {Foxon}, \citenamefont
  {Gallagher}, \citenamefont {Nov\'ak}, \citenamefont {Sinova},\ and\
  \citenamefont {Jungwirth}}]{PhysRevLett.105.227202}%
  \BibitemOpen
  \bibfield  {author} {\bibinfo {author} {\bibfnamefont {J.}~\bibnamefont
  {Ma\ifmmode~\check{s}\else \v{s}\fi{}ek}}, \bibinfo {author} {\bibfnamefont
  {F.}~\bibnamefont {M\'aca}}, \bibinfo {author} {\bibfnamefont
  {J.}~\bibnamefont {Kudrnovsk\'y}}, \bibinfo {author} {\bibfnamefont
  {O.}~\bibnamefont {Makarovsky}}, \bibinfo {author} {\bibfnamefont
  {L.}~\bibnamefont {Eaves}}, \bibinfo {author} {\bibfnamefont {R.~P.}\
  \bibnamefont {Campion}}, \bibinfo {author} {\bibfnamefont {K.~W.}\
  \bibnamefont {Edmonds}}, \bibinfo {author} {\bibfnamefont {A.~W.}\
  \bibnamefont {Rushforth}}, \bibinfo {author} {\bibfnamefont {C.~T.}\
  \bibnamefont {Foxon}}, \bibinfo {author} {\bibfnamefont {B.~L.}\ \bibnamefont
  {Gallagher}}, \bibinfo {author} {\bibfnamefont {V.}~\bibnamefont {Nov\'ak}},
  \bibinfo {author} {\bibfnamefont {J.}~\bibnamefont {Sinova}}, \ and\ \bibinfo
  {author} {\bibfnamefont {T.}~\bibnamefont {Jungwirth}},\ }\href {\doibase
  10.1103/PhysRevLett.105.227202} {\bibfield  {journal} {\bibinfo  {journal}
  {Phys. Rev. Lett.}\ }\textbf {\bibinfo {volume} {105}},\ \bibinfo {pages}
  {227202} (\bibinfo {year} {2010})}\BibitemShut {NoStop}%
\bibitem [{\citenamefont {Bozkurt}\ \emph {et~al.}(2013)\citenamefont
  {Bozkurt}, \citenamefont {Mahani}, \citenamefont {Studer}, \citenamefont
  {Tang}, \citenamefont {Schofield}, \citenamefont {Curson}, \citenamefont
  {Flatt\'e}, \citenamefont {Silov}, \citenamefont {Hirjibehedin},
  \citenamefont {Canali},\ and\ \citenamefont {Koenraad}}]{mc_MF_2013}%
  \BibitemOpen
  \bibfield  {author} {\bibinfo {author} {\bibfnamefont {M.}~\bibnamefont
  {Bozkurt}}, \bibinfo {author} {\bibfnamefont {M.~R.}\ \bibnamefont {Mahani}},
  \bibinfo {author} {\bibfnamefont {P.}~\bibnamefont {Studer}}, \bibinfo
  {author} {\bibfnamefont {J.-M.}\ \bibnamefont {Tang}}, \bibinfo {author}
  {\bibfnamefont {S.~R.}\ \bibnamefont {Schofield}}, \bibinfo {author}
  {\bibfnamefont {N.~J.}\ \bibnamefont {Curson}}, \bibinfo {author}
  {\bibfnamefont {M.~E.}\ \bibnamefont {Flatt\'e}}, \bibinfo {author}
  {\bibfnamefont {A.~Y.}\ \bibnamefont {Silov}}, \bibinfo {author}
  {\bibfnamefont {C.~F.}\ \bibnamefont {Hirjibehedin}}, \bibinfo {author}
  {\bibfnamefont {C.~M.}\ \bibnamefont {Canali}}, \ and\ \bibinfo {author}
  {\bibfnamefont {P.~M.}\ \bibnamefont {Koenraad}},\ }\href {\doibase
  10.1103/PhysRevB.88.205203} {\bibfield  {journal} {\bibinfo  {journal} {Phys.
  Rev. B}\ }\textbf {\bibinfo {volume} {88}},\ \bibinfo {pages} {205203}
  (\bibinfo {year} {2013})}\BibitemShut {NoStop}%
\bibitem [{\citenamefont {Garleff}\ \emph {et~al.}(2008)\citenamefont
  {Garleff}, \citenamefont {\ifmmode~\mbox{\c{C}}\else \c{C}\fi{}elebi},
  \citenamefont {Van~Roy}, \citenamefont {Tang}, \citenamefont {Flatt\'e},\
  and\ \citenamefont {Koenraad}}]{koenraad_prb08}%
  \BibitemOpen
  \bibfield  {author} {\bibinfo {author} {\bibfnamefont {J.~K.}\ \bibnamefont
  {Garleff}}, \bibinfo {author} {\bibfnamefont {C.}~\bibnamefont
  {\ifmmode~\mbox{\c{C}}\else \c{C}\fi{}elebi}}, \bibinfo {author}
  {\bibfnamefont {W.}~\bibnamefont {Van~Roy}}, \bibinfo {author} {\bibfnamefont
  {J.-M.}\ \bibnamefont {Tang}}, \bibinfo {author} {\bibfnamefont {M.~E.}\
  \bibnamefont {Flatt\'e}}, \ and\ \bibinfo {author} {\bibfnamefont {P.~M.}\
  \bibnamefont {Koenraad}},\ }\href {\doibase 10.1103/PhysRevB.78.075313}
  {\bibfield  {journal} {\bibinfo  {journal} {Phys. Rev. B}\ }\textbf {\bibinfo
  {volume} {78}},\ \bibinfo {pages} {075313} (\bibinfo {year}
  {2008})}\BibitemShut {NoStop}%
\bibitem [{\citenamefont {Richardella}\ \emph {et~al.}(2009)\citenamefont
  {Richardella}, \citenamefont {Kitchen},\ and\ \citenamefont
  {Yazdani}}]{yazdani_prb09}%
  \BibitemOpen
  \bibfield  {author} {\bibinfo {author} {\bibfnamefont {A.}~\bibnamefont
  {Richardella}}, \bibinfo {author} {\bibfnamefont {D.}~\bibnamefont
  {Kitchen}}, \ and\ \bibinfo {author} {\bibfnamefont {A.}~\bibnamefont
  {Yazdani}},\ }\href@noop {} {\bibfield  {journal} {\bibinfo  {journal} {Phys.
  Rev. B}\ }\textbf {\bibinfo {volume} {80}},\ \bibinfo {pages} {045318}
  (\bibinfo {year} {2009})}\BibitemShut {NoStop}%
\bibitem [{\citenamefont {Bocquel}\ \emph {et~al.}(2013)\citenamefont
  {Bocquel}, \citenamefont {Kortan}, \citenamefont {Sahin}, \citenamefont
  {Campion}, \citenamefont {Gallagher}, \citenamefont {Flatt{\'e}},\ and\
  \citenamefont {Koenraad}}]{bocq_prb13}%
  \BibitemOpen
  \bibfield  {author} {\bibinfo {author} {\bibfnamefont {J.}~\bibnamefont
  {Bocquel}}, \bibinfo {author} {\bibfnamefont {V.~R.}\ \bibnamefont {Kortan}},
  \bibinfo {author} {\bibfnamefont {C.}~\bibnamefont {Sahin}}, \bibinfo
  {author} {\bibfnamefont {R.~P.}\ \bibnamefont {Campion}}, \bibinfo {author}
  {\bibfnamefont {B.~L.}\ \bibnamefont {Gallagher}}, \bibinfo {author}
  {\bibfnamefont {M.~E.}\ \bibnamefont {Flatt{\'e}}}, \ and\ \bibinfo {author}
  {\bibfnamefont {P.~M.}\ \bibnamefont {Koenraad}},\ }\href@noop {} {\bibfield
  {journal} {\bibinfo  {journal} {Phys. Rev. B}\ }\textbf {\bibinfo {volume}
  {87}},\ \bibinfo {pages} {075421} (\bibinfo {year} {2013})}\BibitemShut
  {NoStop}%
\bibitem [{\citenamefont {M{\"u}hlenberend}\ \emph {et~al.}(2013)\citenamefont
  {M{\"u}hlenberend}, \citenamefont {Gruyters},\ and\ \citenamefont
  {Berndt}}]{svenja_PRB_13}%
  \BibitemOpen
  \bibfield  {author} {\bibinfo {author} {\bibfnamefont {S.}~\bibnamefont
  {M{\"u}hlenberend}}, \bibinfo {author} {\bibfnamefont {M.}~\bibnamefont
  {Gruyters}}, \ and\ \bibinfo {author} {\bibfnamefont {R.}~\bibnamefont
  {Berndt}},\ }\href@noop {} {\bibfield  {journal} {\bibinfo  {journal} {Phys.
  Rev. B}\ }\textbf {\bibinfo {volume} {88}},\ \bibinfo {pages} {115301}
  (\bibinfo {year} {2013})}\BibitemShut {NoStop}%
\bibitem [{\citenamefont {Malguth1}\ \emph {et~al.}(2008)\citenamefont
  {Malguth1}, \citenamefont {Hoffmann},\ and\ \citenamefont
  {Phillips}}]{Malguth_phys_status}%
  \BibitemOpen
  \bibfield  {author} {\bibinfo {author} {\bibfnamefont {E.}~\bibnamefont
  {Malguth1}}, \bibinfo {author} {\bibfnamefont {A.}~\bibnamefont {Hoffmann}},
  \ and\ \bibinfo {author} {\bibfnamefont {M.~R.}\ \bibnamefont {Phillips}},\
  }\href@noop {} {\bibfield  {journal} {\bibinfo  {journal} {physica status
  solidi (b)}\ }\textbf {\bibinfo {volume} {245}},\ \bibinfo {pages} {455}
  (\bibinfo {year} {2008})}\BibitemShut {NoStop}%
\bibitem [{\citenamefont {Chadi}(1977)}]{chadi_prb77}%
  \BibitemOpen
  \bibfield  {author} {\bibinfo {author} {\bibfnamefont {D.~J.}\ \bibnamefont
  {Chadi}},\ }\href@noop {} {\bibfield  {journal} {\bibinfo  {journal} {Phys.
  Rev. B}\ }\textbf {\bibinfo {volume} {16}},\ \bibinfo {pages} {790} (\bibinfo
  {year} {1977})}\BibitemShut {NoStop}%
\bibitem [{\citenamefont {Slater}\ and\ \citenamefont
  {Koster}(1954)}]{slaterkoster_pr54}%
  \BibitemOpen
  \bibfield  {author} {\bibinfo {author} {\bibfnamefont {J.~C.}\ \bibnamefont
  {Slater}}\ and\ \bibinfo {author} {\bibfnamefont {G.~F.}\ \bibnamefont
  {Koster}},\ }\href@noop {} {\bibfield  {journal} {\bibinfo  {journal} {Phys.
  Rev.}\ }\textbf {\bibinfo {volume} {94}},\ \bibinfo {pages} {1498} (\bibinfo
  {year} {1954})}\BibitemShut {NoStop}%
\bibitem [{\citenamefont {Papaconstantopoulos}\ and\ \citenamefont
  {Mehl}(2003)}]{papac_jpcm03}%
  \BibitemOpen
  \bibfield  {author} {\bibinfo {author} {\bibfnamefont {D.~A.}\ \bibnamefont
  {Papaconstantopoulos}}\ and\ \bibinfo {author} {\bibfnamefont {M.~J.}\
  \bibnamefont {Mehl}},\ }\href@noop {} {\bibfield  {journal} {\bibinfo
  {journal} {J. Phys.: Cond. Mat.}\ }\textbf {\bibinfo {volume} {15}},\
  \bibinfo {pages} {R413} (\bibinfo {year} {2003})}\BibitemShut {NoStop}%
\bibitem [{\citenamefont {Jancu}\ \emph {et~al.}(1998)\citenamefont {Jancu},
  \citenamefont {Scholz}, \citenamefont {Beltram},\ and\ \citenamefont
  {Bassani}}]{Bassani_PRB_57_6493}%
  \BibitemOpen
  \bibfield  {author} {\bibinfo {author} {\bibfnamefont {J.-M.}\ \bibnamefont
  {Jancu}}, \bibinfo {author} {\bibfnamefont {R.}~\bibnamefont {Scholz}},
  \bibinfo {author} {\bibfnamefont {F.}~\bibnamefont {Beltram}}, \ and\
  \bibinfo {author} {\bibfnamefont {F.}~\bibnamefont {Bassani}},\ }\href@noop
  {} {\bibfield  {journal} {\bibinfo  {journal} {Phys. Rev. B}\ }\textbf
  {\bibinfo {volume} {57}},\ \bibinfo {pages} {6493} (\bibinfo {year}
  {1998})}\BibitemShut {NoStop}%
\bibitem [{\citenamefont {Chadi}(1978)}]{chadi_prl78}%
  \BibitemOpen
  \bibfield  {author} {\bibinfo {author} {\bibfnamefont {D.~J.}\ \bibnamefont
  {Chadi}},\ }\href@noop {} {\bibfield  {journal} {\bibinfo  {journal} {Phys.
  Rev. Lett.}\ }\textbf {\bibinfo {volume} {41}},\ \bibinfo {pages} {1062}
  (\bibinfo {year} {1978})}\BibitemShut {NoStop}%
\bibitem [{\citenamefont {Chadi}(1979)}]{chadi_prb79}%
  \BibitemOpen
  \bibfield  {author} {\bibinfo {author} {\bibfnamefont {D.~J.}\ \bibnamefont
  {Chadi}},\ }\href@noop {} {\bibfield  {journal} {\bibinfo  {journal} {Phys.
  Rev. B}\ }\textbf {\bibinfo {volume} {19}},\ \bibinfo {pages} {2074}
  (\bibinfo {year} {1979})}\BibitemShut {NoStop}%
\bibitem [{\citenamefont {\ifmmode~\mbox{\c{C}}\else \c{C}\fi{}elebi}\ \emph
  {et~al.}(2010)\citenamefont {\ifmmode~\mbox{\c{C}}\else \c{C}\fi{}elebi},
  \citenamefont {Garleff}, \citenamefont {Silov}, \citenamefont {Yakunin},
  \citenamefont {Koenraad}, \citenamefont {Van~Roy}, \citenamefont {Tang},\
  and\ \citenamefont {Flatt\'e}}]{celebi_prl10}%
  \BibitemOpen
  \bibfield  {author} {\bibinfo {author} {\bibfnamefont {C.}~\bibnamefont
  {\ifmmode~\mbox{\c{C}}\else \c{C}\fi{}elebi}}, \bibinfo {author}
  {\bibfnamefont {J.~K.}\ \bibnamefont {Garleff}}, \bibinfo {author}
  {\bibfnamefont {A.~Y.}\ \bibnamefont {Silov}}, \bibinfo {author}
  {\bibfnamefont {A.~M.}\ \bibnamefont {Yakunin}}, \bibinfo {author}
  {\bibfnamefont {P.~M.}\ \bibnamefont {Koenraad}}, \bibinfo {author}
  {\bibfnamefont {W.}~\bibnamefont {Van~Roy}}, \bibinfo {author} {\bibfnamefont
  {J.-M.}\ \bibnamefont {Tang}}, \ and\ \bibinfo {author} {\bibfnamefont
  {M.~E.}\ \bibnamefont {Flatt\'e}},\ }\href {\doibase
  10.1103/PhysRevLett.104.086404} {\bibfield  {journal} {\bibinfo  {journal}
  {Phys. Rev. Lett.}\ }\textbf {\bibinfo {volume} {104}},\ \bibinfo {pages}
  {086404} (\bibinfo {year} {2010})}\BibitemShut {NoStop}%
\bibitem [{\citenamefont {Yakunin}\ \emph {et~al.}(2007)\citenamefont
  {Yakunin}, \citenamefont {Silov}, \citenamefont {Koenraad}, \citenamefont
  {Tang}, \citenamefont {Flatt{\'e}}, \citenamefont {Primus}, \citenamefont
  {Roy}, \citenamefont {Boeck}, \citenamefont {Monakhov}, \citenamefont
  {Romanov}, \citenamefont {Panaiotti},\ and\ \citenamefont
  {Averkiev}}]{paul_nature_2007}%
  \BibitemOpen
  \bibfield  {author} {\bibinfo {author} {\bibfnamefont {A.~M.}\ \bibnamefont
  {Yakunin}}, \bibinfo {author} {\bibfnamefont {A.~Y.}\ \bibnamefont {Silov}},
  \bibinfo {author} {\bibfnamefont {P.~M.}\ \bibnamefont {Koenraad}}, \bibinfo
  {author} {\bibfnamefont {J.-M.}\ \bibnamefont {Tang}}, \bibinfo {author}
  {\bibfnamefont {M.~E.}\ \bibnamefont {Flatt{\'e}}}, \bibinfo {author}
  {\bibfnamefont {J.-L.}\ \bibnamefont {Primus}}, \bibinfo {author}
  {\bibfnamefont {W.~V.}\ \bibnamefont {Roy}}, \bibinfo {author} {\bibfnamefont
  {J.~D.}\ \bibnamefont {Boeck}}, \bibinfo {author} {\bibfnamefont {A.~M.}\
  \bibnamefont {Monakhov}}, \bibinfo {author} {\bibfnamefont {K.~S.}\
  \bibnamefont {Romanov}}, \bibinfo {author} {\bibfnamefont {I.~E.}\
  \bibnamefont {Panaiotti}}, \ and\ \bibinfo {author} {\bibfnamefont {N.~S.}\
  \bibnamefont {Averkiev}},\ }\href@noop {} {\bibfield  {journal} {\bibinfo
  {journal} {Nat. Mater.}\ }\textbf {\bibinfo {volume} {6}},\ \bibinfo {pages}
  {512} (\bibinfo {year} {2007})}\BibitemShut {NoStop}%
\bibitem [{\citenamefont {Niquet}\ \emph {et~al.}(2009)\citenamefont {Niquet},
  \citenamefont {Rideau}, \citenamefont {Tavernier}, \citenamefont {Jaouen},\
  and\ \citenamefont {Blase}}]{PhysRevB.79.245201}%
  \BibitemOpen
  \bibfield  {author} {\bibinfo {author} {\bibfnamefont {Y.~M.}\ \bibnamefont
  {Niquet}}, \bibinfo {author} {\bibfnamefont {D.}~\bibnamefont {Rideau}},
  \bibinfo {author} {\bibfnamefont {C.}~\bibnamefont {Tavernier}}, \bibinfo
  {author} {\bibfnamefont {H.}~\bibnamefont {Jaouen}}, \ and\ \bibinfo {author}
  {\bibfnamefont {X.}~\bibnamefont {Blase}},\ }\href {\doibase
  10.1103/PhysRevB.79.245201} {\bibfield  {journal} {\bibinfo  {journal} {Phys.
  Rev. B}\ }\textbf {\bibinfo {volume} {79}},\ \bibinfo {pages} {245201}
  (\bibinfo {year} {2009})}\BibitemShut {NoStop}%
\bibitem [{\citenamefont {Zieli\ifmmode~\acute{n}\else
  \'{n}\fi{}ski}(2012)}]{PhysRevB.86.115424}%
  \BibitemOpen
  \bibfield  {author} {\bibinfo {author} {\bibfnamefont {M.}~\bibnamefont
  {Zieli\ifmmode~\acute{n}\else \'{n}\fi{}ski}},\ }\href {\doibase
  10.1103/PhysRevB.86.115424} {\bibfield  {journal} {\bibinfo  {journal} {Phys.
  Rev. B}\ }\textbf {\bibinfo {volume} {86}},\ \bibinfo {pages} {115424}
  (\bibinfo {year} {2012})}\BibitemShut {NoStop}%
\bibitem [{\citenamefont {Blaha}\ \emph {et~al.}(2001)\citenamefont {Blaha},
  \citenamefont {Schwarz}, \citenamefont {Madsen}, \citenamefont {Kvasnicka},\
  and\ \citenamefont {Luitz}}]{Wien2k_package}%
  \BibitemOpen
  \bibfield  {author} {\bibinfo {author} {\bibfnamefont {P.}~\bibnamefont
  {Blaha}}, \bibinfo {author} {\bibfnamefont {K.}~\bibnamefont {Schwarz}},
  \bibinfo {author} {\bibfnamefont {G.~K.~H.}\ \bibnamefont {Madsen}}, \bibinfo
  {author} {\bibfnamefont {D.}~\bibnamefont {Kvasnicka}}, \ and\ \bibinfo
  {author} {\bibfnamefont {J.}~\bibnamefont {Luitz}},\ }\href@noop {} {\emph
  {\bibinfo {title} {WIEN2k, An Augmented Plane Wave Plus Local Orbitals
  Program for Calculating Crystal properties (Vienna University of Technology,
  Austria)}}} (\bibinfo {year} {2001})\BibitemShut {NoStop}%
\bibitem [{\citenamefont {Perdew}\ \emph {et~al.}(1996)\citenamefont {Perdew},
  \citenamefont {Burke},\ and\ \citenamefont {Ernzerhof}}]{perdew96}%
  \BibitemOpen
  \bibfield  {author} {\bibinfo {author} {\bibfnamefont {J.~P.}\ \bibnamefont
  {Perdew}}, \bibinfo {author} {\bibfnamefont {K.}~\bibnamefont {Burke}}, \
  and\ \bibinfo {author} {\bibfnamefont {M.}~\bibnamefont {Ernzerhof}},\ }\href
  {\doibase 10.1103/PhysRevLett.77.3865} {\bibfield  {journal} {\bibinfo
  {journal} {Phys. Rev. Lett.}\ }\textbf {\bibinfo {volume} {77}},\ \bibinfo
  {pages} {3865} (\bibinfo {year} {1996})}\BibitemShut {NoStop}%
\bibitem [{\citenamefont {Soler}\ \emph {et~al.}(2002)\citenamefont {Soler},
  \citenamefont {Artacho}, \citenamefont {Gale}, \citenamefont {García},
  \citenamefont {Junquera}, \citenamefont {Ordejón},\ and\ \citenamefont
  {Sánchez-Portal}}]{siesta}%
  \BibitemOpen
  \bibfield  {author} {\bibinfo {author} {\bibfnamefont {J.~M.}\ \bibnamefont
  {Soler}}, \bibinfo {author} {\bibfnamefont {E.}~\bibnamefont {Artacho}},
  \bibinfo {author} {\bibfnamefont {J.~D.}\ \bibnamefont {Gale}}, \bibinfo
  {author} {\bibfnamefont {A.}~\bibnamefont {García}}, \bibinfo {author}
  {\bibfnamefont {J.}~\bibnamefont {Junquera}}, \bibinfo {author}
  {\bibfnamefont {P.}~\bibnamefont {Ordejón}}, \ and\ \bibinfo {author}
  {\bibfnamefont {D.}~\bibnamefont {Sánchez-Portal}},\ }\href
  {http://stacks.iop.org/0953-8984/14/i=11/a=302} {\bibfield  {journal}
  {\bibinfo  {journal} {Journal of Physics: Condensed Matter}\ }\textbf
  {\bibinfo {volume} {14}},\ \bibinfo {pages} {2745} (\bibinfo {year}
  {2002})}\BibitemShut {NoStop}%
\bibitem [{\citenamefont {Anisimov}\ \emph {et~al.}(1991)\citenamefont
  {Anisimov}, \citenamefont {Zaanen},\ and\ \citenamefont
  {Andersen}}]{anisimov1991band}%
  \BibitemOpen
  \bibfield  {author} {\bibinfo {author} {\bibfnamefont {V.~I.}\ \bibnamefont
  {Anisimov}}, \bibinfo {author} {\bibfnamefont {J.}~\bibnamefont {Zaanen}}, \
  and\ \bibinfo {author} {\bibfnamefont {O.~K.}\ \bibnamefont {Andersen}},\
  }\href@noop {} {\bibfield  {journal} {\bibinfo  {journal} {Phys. Rev. B}\
  }\textbf {\bibinfo {volume} {44}},\ \bibinfo {pages} {943} (\bibinfo {year}
  {1991})}\BibitemShut {NoStop}%
\bibitem [{\citenamefont {Sato}\ \emph {et~al.}(2010)\citenamefont {Sato},
  \citenamefont {Bergqvist}, \citenamefont {Kudrnovsk{\'y}}, \citenamefont
  {Dederichs}, \citenamefont {Eriksson}, \citenamefont {Turek}, \citenamefont
  {Sanyal}, \citenamefont {Bouzerar}, \citenamefont {Katayama-Yoshida},
  \citenamefont {Dinh}, \citenamefont {Fukushima}, \citenamefont {Kizaki},\
  and\ \citenamefont {Zeller}}]{uppsala_DMSreview_2010}%
  \BibitemOpen
  \bibfield  {author} {\bibinfo {author} {\bibfnamefont {K.}~\bibnamefont
  {Sato}}, \bibinfo {author} {\bibfnamefont {L.}~\bibnamefont {Bergqvist}},
  \bibinfo {author} {\bibfnamefont {J.}~\bibnamefont {Kudrnovsk{\'y}}},
  \bibinfo {author} {\bibfnamefont {P.~H.}\ \bibnamefont {Dederichs}}, \bibinfo
  {author} {\bibfnamefont {O.}~\bibnamefont {Eriksson}}, \bibinfo {author}
  {\bibfnamefont {I.}~\bibnamefont {Turek}}, \bibinfo {author} {\bibfnamefont
  {B.}~\bibnamefont {Sanyal}}, \bibinfo {author} {\bibfnamefont
  {G.}~\bibnamefont {Bouzerar}}, \bibinfo {author} {\bibfnamefont
  {H.}~\bibnamefont {Katayama-Yoshida}}, \bibinfo {author} {\bibfnamefont
  {V.~A.}\ \bibnamefont {Dinh}}, \bibinfo {author} {\bibfnamefont
  {T.}~\bibnamefont {Fukushima}}, \bibinfo {author} {\bibfnamefont
  {H.}~\bibnamefont {Kizaki}}, \ and\ \bibinfo {author} {\bibfnamefont
  {R.}~\bibnamefont {Zeller}},\ }\href@noop {} {\bibfield  {journal} {\bibinfo
  {journal} {Rev. Mod. Phys.}\ }\textbf {\bibinfo {volume} {82}},\ \bibinfo
  {pages} {1633} (\bibinfo {year} {2010})}\BibitemShut {NoStop}%
\bibitem [{\citenamefont {Belhadji}\ \emph {et~al.}(2007)\citenamefont
  {Belhadji}, \citenamefont {Bergqvist}, \citenamefont {Zeller}, \citenamefont
  {Dederichs}, \citenamefont {Sato},\ and\ \citenamefont
  {Katayama-Yoshida}}]{belhadji2007trends}%
  \BibitemOpen
  \bibfield  {author} {\bibinfo {author} {\bibfnamefont {B.}~\bibnamefont
  {Belhadji}}, \bibinfo {author} {\bibfnamefont {L.}~\bibnamefont {Bergqvist}},
  \bibinfo {author} {\bibfnamefont {R.}~\bibnamefont {Zeller}}, \bibinfo
  {author} {\bibfnamefont {P.~H.}\ \bibnamefont {Dederichs}}, \bibinfo {author}
  {\bibfnamefont {K.}~\bibnamefont {Sato}}, \ and\ \bibinfo {author}
  {\bibfnamefont {H.}~\bibnamefont {Katayama-Yoshida}},\ }\href@noop {}
  {\bibfield  {journal} {\bibinfo  {journal} {Journal of Physics: Condensed
  Matter}\ }\textbf {\bibinfo {volume} {19}},\ \bibinfo {pages} {436227}
  (\bibinfo {year} {2007})}\BibitemShut {NoStop}%
\bibitem [{\citenamefont {Gopal}\ and\ \citenamefont
  {Spaldin}(2006)}]{gopal2006magnetic}%
  \BibitemOpen
  \bibfield  {author} {\bibinfo {author} {\bibfnamefont {P.}~\bibnamefont
  {Gopal}}\ and\ \bibinfo {author} {\bibfnamefont {N.~A.}\ \bibnamefont
  {Spaldin}},\ }\href@noop {} {\bibfield  {journal} {\bibinfo  {journal} {Phys.
  Rev. B}\ }\textbf {\bibinfo {volume} {74}},\ \bibinfo {pages} {094418}
  (\bibinfo {year} {2006})}\BibitemShut {NoStop}%
\bibitem [{\citenamefont {Schairer}\ and\ \citenamefont
  {Schmidt}(1974)}]{schairer_prb74}%
  \BibitemOpen
  \bibfield  {author} {\bibinfo {author} {\bibfnamefont {W.}~\bibnamefont
  {Schairer}}\ and\ \bibinfo {author} {\bibfnamefont {M.}~\bibnamefont
  {Schmidt}},\ }\href@noop {} {\bibfield  {journal} {\bibinfo  {journal} {Phys.
  Rev. B}\ }\textbf {\bibinfo {volume} {10}},\ \bibinfo {pages} {2501}
  (\bibinfo {year} {1974})}\BibitemShut {NoStop}%
\bibitem [{\citenamefont {Lee}\ and\ \citenamefont
  {Anderson}(1964)}]{lee_ssc64}%
  \BibitemOpen
  \bibfield  {author} {\bibinfo {author} {\bibfnamefont {T.}~\bibnamefont
  {Lee}}\ and\ \bibinfo {author} {\bibfnamefont {W.~W.}\ \bibnamefont
  {Anderson}},\ }\href@noop {} {\bibfield  {journal} {\bibinfo  {journal}
  {Solid State Commun.}\ }\textbf {\bibinfo {volume} {2}},\ \bibinfo {pages}
  {265} (\bibinfo {year} {1964})}\BibitemShut {NoStop}%
\bibitem [{\citenamefont {Chapman}\ and\ \citenamefont
  {Hutchinson}(1967)}]{chapman_prl67}%
  \BibitemOpen
  \bibfield  {author} {\bibinfo {author} {\bibfnamefont {R.~A.}\ \bibnamefont
  {Chapman}}\ and\ \bibinfo {author} {\bibfnamefont {W.~G.}\ \bibnamefont
  {Hutchinson}},\ }\href@noop {} {\bibfield  {journal} {\bibinfo  {journal}
  {Phys. Rev. Lett.}\ }\textbf {\bibinfo {volume} {18}},\ \bibinfo {pages}
  {443} (\bibinfo {year} {1967})}\BibitemShut {NoStop}%
\bibitem [{\citenamefont {Linnarsson}\ \emph {et~al.}(1997)\citenamefont
  {Linnarsson}, \citenamefont {Janzen}, \citenamefont {Monemar}, \citenamefont
  {Kleverman},\ and\ \citenamefont {Thilderkvist}}]{linnarsson_prb97}%
  \BibitemOpen
  \bibfield  {author} {\bibinfo {author} {\bibfnamefont {M.}~\bibnamefont
  {Linnarsson}}, \bibinfo {author} {\bibfnamefont {E.}~\bibnamefont {Janzen}},
  \bibinfo {author} {\bibfnamefont {B.}~\bibnamefont {Monemar}}, \bibinfo
  {author} {\bibfnamefont {M.}~\bibnamefont {Kleverman}}, \ and\ \bibinfo
  {author} {\bibfnamefont {A.}~\bibnamefont {Thilderkvist}},\ }\href@noop {}
  {\bibfield  {journal} {\bibinfo  {journal} {Phys. Rev. B}\ }\textbf {\bibinfo
  {volume} {55}},\ \bibinfo {pages} {6938} (\bibinfo {year}
  {1997})}\BibitemShut {NoStop}%
\bibitem [{\citenamefont {Mahani}\ \emph {et~al.}(2014)\citenamefont {Mahani},
  \citenamefont {Pertsova},\ and\ \citenamefont {Canali}}]{rm_JPCM}%
  \BibitemOpen
  \bibfield  {author} {\bibinfo {author} {\bibfnamefont {M.}~\bibnamefont
  {Mahani}}, \bibinfo {author} {\bibfnamefont {A.}~\bibnamefont {Pertsova}}, \
  and\ \bibinfo {author} {\bibfnamefont {C.}~\bibnamefont {Canali}},\
  }\href@noop {} {\bibfield  {journal} {\bibinfo  {journal} {arXiv:1402.3069}\
  } (\bibinfo {year} {2014})}\BibitemShut {NoStop}%
\end{thebibliography}%
\end{document}